\documentclass[a4paper,11pt]{article}
\pdfoutput=1
\usepackage{jheppub,undertilde} 
\usepackage[T1]{fontenc} 
\usepackage{amsmath,amssymb,graphicx,bbm,mathrsfs,nicefrac}
\usepackage{graphicx}
\usepackage{feynmf}
\usepackage{tikz}
\usepackage{multirow}
\usepackage{tabularx}
\usepackage{bbm}
\usepackage{slashed,color}

\newcommand\sss{\scriptscriptstyle}

\usepackage{colortbl,framed,graphicx,subcaption}

 \def\lra#1{\overset{\text{\scriptsize$\leftrightarrow$}}{#1}}

\providecommand{\tabularnewline}{\\}

\def\bsp#1\esp{\begin{split}#1\end{split}}

\newcommand{\be}{\begin{equation}} 
\newcommand{\ee}{\end{equation}}  
\newcommand{\bea}{\begin{eqnarray}}  
\newcommand{\eea}{\end{eqnarray}}  
\def\bpm{\begin{pmatrix}}
\def\epm{\end{pmatrix}}

\def\bsp#1\esp{\begin{split}#1\end{split}}
\def\spa#1.#2{\left\langle #1 \, #2 \right\rangle}
\def\spb#1.#2{\left[ #1 \, #2 \right]}
\def\spab#1.#2.#3{\left\langle #1 |#2| #3 \right]}
\def\spaa#1.#2.#3.#4{\left\langle #1 |#2 |#3 | #4 \right\rangle}
\def\spbb#1.#2.#3.#4{\left[ #1 | #2 | #3 | #4 \right]}

\newcommand\lsim{\mathrel{\rlap{\lower4pt\hbox{\hskip1pt$\sim$}}
    \raise1pt\hbox{$<$}}}
\newcommand\gsim{\mathrel{\rlap{\lower4pt\hbox{\hskip1pt$\sim$}}
    \raise1pt\hbox{$>$}}}

\newcommand{\mW}{m_{W}}
\newcommand{\sW}{s_{W}}
\newcommand{\cW}{c_{W}}

\providecommand{\tabularnewline}{\\}

\newcommand{\captionfonts}{\small}

\newcommand{\approptoinn}[2]{\mathrel{\vcenter{
  \offinterlineskip\halign{\hfil$##$\cr
    #1\propto\cr\noalign{\kern2pt}#1\sim\cr\noalign{\kern-2pt}}}}}

\makeatletter  
\long\def\@akecaption#1#2{%
  \vskip\abovecaptionskip
  \sbox\@tempboxa{{\captionfonts #1: #2}}%
  \ifdim \wd\@tempboxa >\hsize
    {\captionfonts #1: #2\par}
  \else
    \hbox to\hsize{\hfil\box\@tempboxa\hfil}%
  \fi
  \vskip\belowcaptionskip}
  
\makeatother   

\title{Higher Order QCD predictions for Associated Higgs production with anomalous couplings to gauge bosons}

\author[a]{Ken Mimasu}
\author[a]{Ver\'onica Sanz}
\author[b]{and Ciaran Williams}
 
\affiliation[a]{Department of Physics and Astronomy, University of Sussex, Brighton BN1 9QH, UK}
\affiliation[b]{Department of Physics, University at Buffalo, The State University of New York, Buffalo 14260 USA}
 
\abstract{We present predictions for the associated production of a Higgs boson at NLO+PS accuracy, including the effect of anomalous interactions between the Higgs and gauge bosons. 
We present our results in different frameworks, one in which the interaction vertex between the Higgs boson and Standard Model $W$ and $Z$ bosons is parameterized in terms 
of general Lorentz structures, and one in which Electroweak symmetry breaking is manifestly linear and the resulting operators arise through a six-dimensional effective field theory framework. We present analytic calculations of the Standard Model and Beyond the Standard Model contributions, and discuss the phenomenological impact of the higher order pieces. 
Our results are implemented in the NLO Monte Carlo program MCFM, and interfaced to shower Monte Carlos through the {\sc Powheg} box framework. 
}

\emailAdd{k.mimasu@sussex.ac.uk} 
\emailAdd{v.sanz@sussex.ac.uk}
\emailAdd{ciaranwi@buffalo.edu}


\begin{document}
\maketitle
\flushbottom

\section{Introduction} 

The LHC's discovery of a particle consistent with the predicted Standard Model (SM) Higgs boson has 
opened the door to a full understanding of electroweak symmetry breaking in nature. One of the key aims 
of Run II of the LHC is to study the properties and interactions of the Higgs in as much 
detail as possible, with the ultimate goal of confirming, or seriously constraining, the possibility of new physics 
and/or anomalous interactions. 

One of the most interesting electroweak processes to study at the LHC is the interaction of the Higgs boson 
with massive vector bosons $(W,Z)$.  The primary role of the Higgs is to generate masses for these
particles and ensure perturbative unitarity in vector boson scattering and any deviation from the SM Higgs-vector boson
vertex could be indicative of new physics contributions. 
At the LHC the dominant Higgs production mechanism occurs through the fusion of gluons via a top quark loop. 
Therefore the total inclusive Higgs cross section at the LHC is more sensitive to the top Yukawa coupling than potential 
anomalous interactions of the Higgs with vector bosons. An obvious place to study the interaction between the Higgs 
and vector bosons are the decays $H\rightarrow VV^*$. However since the Higgs is considerably lighter than the $2m_V$ threshold
the decay phase space is restricted, forcing one of the final state vector bosons off-shell. Consequently, anomalous interactions that modify the high energy behaviour of the vertex, are suppressed due to the kinematic requirements. 
Accordingly, the best places to constrain anomalous interactions of the Higgs and vector bosons are those sensitive to said 
vertex in production, namely Vector Boson Fusion (VBF), Higgs in association with a hard jet, and associated production ($VH$). Of these, associated production -- which 
occurs through an $s$-channel production mechanism -- is particularly appealing, since one can directly probe the high energy behaviour 
of the interaction through, for instance, the invariant mass of the Higgs Vector system, $m_{VH}$ . 

A simple way to encode effects of new physics in the Higgs sector is to study Higgs anomalous couplings (HAC)~\cite{HAC}. This parametrization does not rely on assumptions about whether EWSB is linearly or non-linearly realized, as it only relies on the Higgs as a scalar degree of freedom and the preservation of $U(1)_{EM}$, i.e. by saturating all possible Lorentz structures in the vertex with the lowest number of derivatives. This parametrization was successfully used  at LEP in the study of anomalous trilinear gauge couplings~\cite{LEP-TGCs} and adopted in the study of BSM effects in the Higgs couplings. 

An alternative way to describe indirect effects of new physics is to use an Effective Field Theory (EFT) approach. Within this approach, one could assume a linear realization of EWSB with the Higgs  as a doublet of $SU(2)$, and write down all the relevant operators which satisfy $SU(2)_L\times U(1)_Y$~\cite{Buchmuller}. This effective Lagrangian can be written in several equivalent ways which account for the choice of a basis. In this paper we will be using the proposal in Refs.~\cite{continos, HELatLO}. A translation into other choices of basis can be done using, e.g. the tool {\tt Rosetta}~\cite{Rosetta}. Also, one could write an EFT for a non-linear realization of EWSB as in Refs.~\cite{NLEWSB}. In either case, there is a correspondence between the HAC and EFT approaches, see e.g.~\cite{HELatLO}.

In this paper we focus on searching for BSM effects in Higgs production in association with a massive vector boson.  The Higgs associated production process is defined through the following reaction, 
\begin{eqnarray} 
q(p_1)+\overline{q}(p_2) \rightarrow V^* \rightarrow V(p_V) + H (p_H) 
\end{eqnarray} 
Where $V$ represents an electroweak vector boson. In the SM
$V$ is constrained to be either a $W$ or a $Z$, whilst including the higher dimensional 
operators allow for  potential exchange of an off-shell virtual photon. 
The massive bosons are unstable and their decay products are measured 
at collider experiments. Leptonic decays of the vector boson are the cleanest experimentally, 
whilst the decay $H\rightarrow b\overline{b}$ corresponds to the maximal Higgs branching ratio. 
Therefore, unless otherwise stated, the process we study in this paper corresponds to 
\begin{eqnarray} 
q(p_1)+\overline{q}(p_2) \rightarrow V(\rightarrow \ell_1(p_3)+\ell_2(p_4)) + H(\rightarrow b(p_5) +\overline{b}(p_6)) 
\end{eqnarray} 
$\ell_1$ and $\ell_2$ correspond to either two charged leptons ($V=Z$) or a charged lepton and neutrino ($V=W$). 



Associated production of a Higgs boson and a $Z$ includes the following production process 
\begin{eqnarray} 
g(p_1)+g(p_2)\rightarrow Z(\rightarrow \ell^-(p_3)+\ell^+(p_4)) + H(\rightarrow b(p_5)+\overline{b}(p_6)) 
\end{eqnarray}
This process is formally $\mathcal{O}(\alpha_S^2)$ and therefore a consistent treatment in a perturbative 
expansion would first include this piece at NNLO. However, the large gluon flux at the LHC, coupled with the 
boosted topology typical of experimental searches results in a significant contribution from this initial state. 
The SM contribution corresponds to two types of topologies, one in which the top loop radiates 
the $Z$, and one in which the $Z$ is produced at the $HZZ$ vertex. 
In the SM these two terms destructively interfere.

Given its phenomenological relevance significant theoretical effort has been invested in the $VH$ 
processes. The NNLO corrections for the inclusive (i.e. on-shell $VH$) cross section were presented 
in ref.~\cite{Brein:2003wg} based on the previous calculation of corrections to the Drell-Yan process~\cite{Hamberg:1990np}.  The results of ref.~\cite{Brein:2003wg} included the on-shell contributions from the 
$gg$ pieces described in the proceeding paragraphs (a study of the $gg$ pieces at NLO in the heavy top EFT was presented in ref~\cite{Altenkamp:2012sx}). A second type of heavy quark initiated contribution 
arises at NNLO and contains a $q\overline{q}$ pair. The leading contributions were calculated terms of an asymptotic expansion in $m_t^2$ in ref.~\cite{Brein:2011vx}. A fully differential calculation for the Drell-Yan type $WH$ process was presented in~\cite{Ferrera:2011bk}, and was extended to include NLO $H\rightarrow b\overline{b}$ decays in~\cite{Ferrera:2013yga}. A similar calculation for $ZH$, including the $gg$ diagrams was presented in ref~\cite{Ferrera:2014lca} while a complete, fully differential calculation of the NNLO production (Drell-Yan and heavy quark) with decays for $b\overline{b}$ at NLO can be found in ref.~\cite{Campbell:2016jau}. The calculation of the NLO EW corrections for $WH$ were presented in~\cite{Ciccolini:2003jy}. Fully differential predictions to the $H\rightarrow b\overline{b}$ decay were computed at NNLO in~\cite{Anastasiou:2011qx} (the total inclusive $H\rightarrow b\overline{b}$ rate is known to $\mathcal{O}(\alpha_s^4)$~\cite{Baikov:2005rw}).
In addition to the parton level calculations discussed above there has been significant progress matching parton level predictions to parton showers, allowing for full simulation of the LHC collisions. An NLO matched prediction for $VH$ using the {\sc Powheg}~\cite{Frixione:2007vw} framework was first presented in ref.~\cite{Hamilton:2009za} and using the MC@NLO\cite{Frixione:2002ik} setup~\cite{LatundeDada:2009rr}. Studies using merged NLO samples of $VH+0$ and 1 jet were in {\sc Powheg}~\cite{Luisoni:2013cuh} and SHERPA~\cite{Goncalves:2015mfa}. A study for $WH$ including anomalous couplings was presented at parton level in VBFNLO in ref~.\cite{Campanario:2014lza}. EW corrections have been 
implemented in the HAWK Monte Carlo code~\cite{Denner:2011id}, including a study of anomalous $HVV$ interactions at NLO in QCD.

The aim of this paper is to provide a framework to combine the precision predictions described above with the anomalous coupling prediction in a general HAC or EFT framework. We will do this be by calculating the $VH$ processes at NLO including a general parameterization of the $HVV$ vertex. We then interface our results to the POHWEG-BOX~\cite{Frixione:2007vw,Hamilton:2009za,Alioli:2010xd}
 allowing for full event simulation. This paper proceeds as follows, in section~\ref{sec:EFT} we discuss the implementation of the anomalous couplings 
through the EFT Lagrangian. Section~\ref{sec:Calc} presents our calculation in detail and provides the amplitudes for $VH$ production at NLO including the anomalous couplings. Section~\ref{sec:results} includes predictions at fixed order and NLO+PS accuracy and we present phenomenological results for LHC Run II. We draw our conclusions in section~\ref{sec:concs}. 

\section{The Effective Standard Model} \label{sec:EFT}
In this paper we focus on the effects of heavy New Physics in production of a Higgs boson in association with a vector boson.  We are interested, hence, in three-point  functions involving the Higgs and two vector bosons~\cite{HAC,HELatLO} 
\be\bsp
  {\cal L}_{HAC} = &\ 
      -\frac{1}{4} g_{  hzz}^{(1)} Z_{\mu\nu} Z^{\mu\nu} h
    -g_{  hzz}^{(2)} Z_\nu \partial_\mu Z^{\mu\nu} h  + \frac{1}{2} g_{  hzz}^{(3)} Z_\mu Z^{\mu} h
    -\frac{1}{4} \tilde g_{  hzz} Z_{\mu\nu} \tilde Z^{\mu\nu} h 
\\ &\
    -  \frac{1}{2} g_{  hww}^{(1)} W^{\mu \nu} W^\dag_{\mu\nu} h -  \Big[g_{  hww}^{(2)} W^\nu \partial^\mu W^\dag_{\mu\nu} h + {\rm h.c.} \Big] +  g_{  hww}^{(3)} W_\mu W^{\dag \mu} h     -\frac{1}{2} \tilde g_{  hww} W^{\mu\nu} \tilde W^\dag_{\mu\nu} h
    \\ &\
     - \frac{1}{2} g_{haz}^{(1)} Z_{\mu\nu} F^{\mu\nu} h
        - g_{ haz}^{(2)} Z_\nu \partial_\mu F^{\mu\nu} h
    - \frac{1}{2} \tilde g_{haz} Z_{\mu\nu} \tilde F^{\mu\nu} h 
\esp\label{eq:massbasis}\ee
as well as, possibly, couplings of the Higgs to a vector boson and two fermions. These Higgs anomalous couplings (HAC) are a model-independent parametrization which respects the fundamental symmetries of the SM at energies below electroweak symmetry breaking (EWSB), namely Lorentz and $U(1)_{EM}$ invariance, assuming that the Higgs is a neutral, scalar particle. 

The HAC can be related to an Effective Field Theory approach (EFT), where new resonances participating in EWSB are integrated out. The relation between HAC and EFT depends on assumptions of how EWSB occurs, e.g. whether the symmetry is linearly or non-linearly realized. In this paper we will match results in terms of HAC with a linearly realized EFT in which the Higgs $h$ is part of a doublet of $SU(2)_L$. We follow the conventions for EFT operators in ~\cite{continos,HELatLO}, which are based on the work in Ref.~\cite{SILH}.  The relevant part of the Lagrangian is as follows,

\bea 
\begin{split}
    \label{eq:SILH}
  {\cal L}_{\rm EFT} =&\
  \frac{g'^2\ \bar c_{\sss \gamma}}{\mW^2} \Phi^\dag \Phi B_{\mu\nu} B^{\mu\nu}
   +\frac{g_s^2\ \bar c_{\sss g}}{\mW^2} \Phi^\dag \Phi G_{\mu\nu}^a G_a^{\mu\nu}\\
  &\ 
   + \frac{\bar c_{\sss H}}{2 v^2} \partial^\mu\big[\Phi^\dag \Phi\big] \partial_\mu \big[ \Phi^\dagger \Phi \big]
  + \frac{\bar c_{\sss T}}{2 v^2} \big[ \Phi^\dag {\overleftrightarrow{D}}^\mu \Phi \big] \big[ \Phi^\dag {\overleftrightarrow{D}}_\mu \Phi \big] 
  - \frac{\bar c_{\sss 6} \lambda}{v^2} \big[\Phi^\dag \Phi \big]^3 \\
  &\  - \bigg[
     \frac{\bar c_{\sss u}}{v^2} y_u     \Phi^\dag \Phi\ \Phi^\dag\cdot{\bar Q}_L u_R
   + \frac{\bar c_{\sss d}}{v^2} y_d     \Phi^\dag \Phi\ \Phi {\bar Q}_L d_R
   + \frac{\bar c_{\sss l}}{v^2} y_\ell\ \Phi^\dag \Phi\ \Phi {\bar L}_L e_R
   + {\rm h.c.} \bigg] \\
  &\
  + \frac{i g\ \bar c_{\sss W}}{\mW^2} \big[ \Phi^\dag T_{2k} \overleftrightarrow{D}^\mu \Phi \big]  D^\nu  W_{\mu \nu}^k
  + \frac{i g'\ \bar c_{\sss B}}{2 \mW^2} \big[\Phi^\dag \overleftrightarrow{D}^\mu \Phi \big] \partial^\nu  B_{\mu \nu} \\
  &\
  + \frac{2 i g\ \bar c_{\sss HW}}{\mW^2} \big[D^\mu \Phi^\dag T_{2k} D^\nu \Phi\big] W_{\mu \nu}^k
  + \frac{i g'\ \bar c_{\sss HB}}{\mW^2}  \big[D^\mu \Phi^\dag D^\nu \Phi\big] B_{\mu \nu} \\
  & \
    + \frac{i g\ \tilde c_{\sss HW}}{\mW^2}  D^\mu \Phi^\dag T_{2k} D^\nu \Phi {\widetilde W}_{\mu \nu}^k
  + \frac{i g'\ \tilde c_{\sss HB}}{\mW^2} D^\mu \Phi^\dag D^\nu \Phi {\widetilde B}_{\mu \nu}
  + \frac{g'^2\  \tilde c_{\sss \gamma}}{\mW^2} \Phi^\dag \Phi B_{\mu\nu} {\widetilde B}^{\mu\nu}\\
 &\
  +\!  \frac{g_s^2\ \tilde c_{\sss g}}{\mW^2}      \Phi^\dag \Phi G_{\mu\nu}^a {\widetilde G}^{\mu\nu}_a
  \!+\!  \frac{g^3\ \tilde c_{\sss 3W}}{\mW^2} \epsilon_{ijk} W_{\mu\nu}^i W^\nu{}^j_\rho {\widetilde W}^{\rho\mu k}
  \!+\!  \frac{g_s^3\ \tilde c_{\sss 3G}}{\mW^2} f_{abc} G_{\mu\nu}^a G^\nu{}^b_\rho {\widetilde G}^{\rho\mu c} \ ,
\end{split}
\eea
where $\Phi$ is the Higgs doublet,
\bea
  \Phi = \bpm -i G^+ \\ \frac{1}{\sqrt{2}} \Big[ v + h + i G^0\Big] \epm  \ ,
\eea
and the dual field strength tensors are defined by
\bea
  \widetilde B_{\mu\nu} = \frac12 \epsilon_{\mu\nu\rho\sigma} B^{\rho\sigma} \ , \quad
  \widetilde W_{\mu\nu}^k = \frac12 \epsilon_{\mu\nu\rho\sigma} W^{\rho\sigma k} \ , \quad
  \widetilde G_{\mu\nu}^a = \frac12 \epsilon_{\mu\nu\rho\sigma} G^{\rho\sigma a} \ .
\eea

\begin{table}
  \center
  \begin{tabular}{l |  l}
  \hline\hline \\
    $g^{(1)}_{hzz}$ &  $\frac{2 g}{\cW^2 \mW} \Big[ \bar c_{  HB} \sW^2 - 4 \bar c_{  \gamma} \sW^4 + \cW^2 \bar c_{  HW}\Big]$ \\[1.2ex]
    $g^{(2)}_{  hzz}$  &
        $\frac{g}{\cW^2 \mW} \Big[(\bar c_{  HW} +\bar c_{  W}) \cW^2  + (\bar c_{  B} + \bar c_{  HB}) \sW^2 \Big]$\\[1.2ex]
    $g^{(3)}_{  hzz}$   & $\frac{g \mW}{\cW^2} \Big[ 1 -\frac12 \bar c_{  H} - 2 \bar c_{  T}
          +8 \bar c_{ \gamma} \frac{\sW^4}{\cW^2}  \Big]$ \\[1.2ex]
            $\tilde g_{  hzz}$ & 
       $\frac{2 g}{\cW^2 \mW} \Big[ \tilde c_{  HB} \sW^2 - 4 \tilde c_{  \gamma} \sW^4 + \cW^2 \tilde c_{  HW}\Big]$ \\[1.2ex]
    $g^{(1)}_{  haz}$ &
          $\frac{g \sW}{\cW \mW} \Big[  \bar c_{  HW} - \bar c_{  HB} + 8 \bar c_{  \gamma} \sW^2\Big]$ \\[1.2ex]
    $g^{(2)}_{  haz}$   &
      $\frac{g \sW}{\cW \mW} \Big[  \bar c_{  HW} - \bar c_{  HB} - \bar c_{  B} + \bar c_{  W}\Big]$ \\[1.2ex]
         $\tilde g_{  haz}$   &
         $\frac{g \sW}{\cW \mW} \Big[  \tilde c_{  HW} - \tilde c_{  HB} + 8 \tilde c_{  \gamma} \sW^2\Big]$ \\[1.2ex]
    $g^{(1)}_{  hww}$  &  $\frac{2 g}{\mW} \bar c_{  HW}$ \\[1.2ex]
     $g^{(2)}_{  hww}$ &
      $\frac{g}{\mW} \Big[ \bar c_{  W} + \bar c_{  HW} \Big]$ \\ [1.2ex]
    $\tilde g_{  hww}$  & $\frac{2 g}{\mW} \tilde c_{  HW}$ \\ \\
    \hline\hline
    \end{tabular}
  \caption{Translation between the HAC and EFT coefficients.}
  \label{trans}
\end{table}

In Table~\ref{trans} we show the relation between HAC and coefficients of the EFT. 
The basis we have chosen in this paper is not a unique choice and one can use, for instance, {\tt Rosetta}~\cite{Rosetta} as a tool to translate among different basis.
 
\begin{figure}
\begin{center}
\includegraphics[width=12cm]{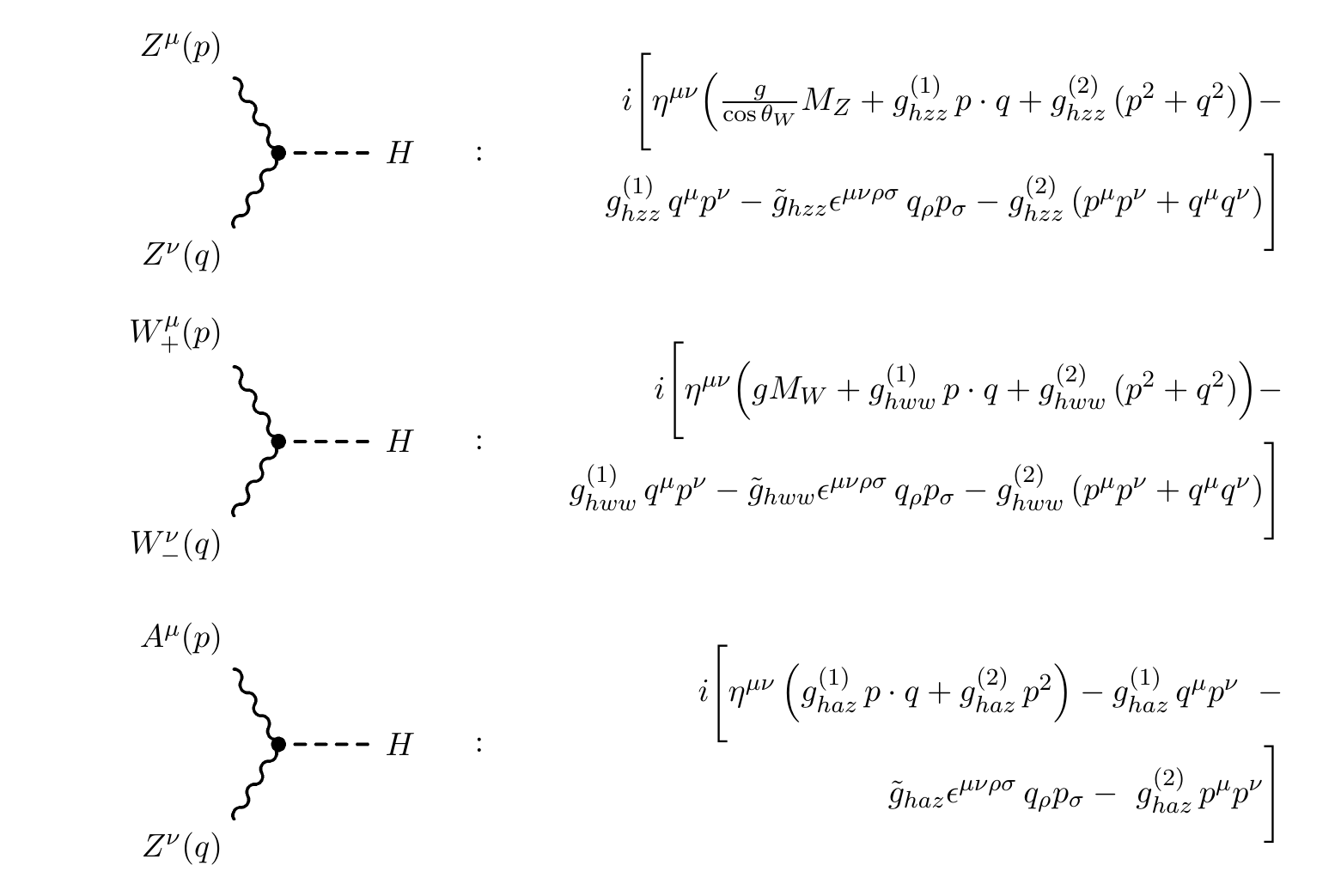}
\end{center}
\caption{\label{f:feynrules} Feynman rules for the Higgs-$Z$-$Z$, Higgs-$W^+$-$W^-$ and Higgs-$\gamma$-$Z$ vertices in the Higgs Anomalous Coupling description of eq.~\ref{eq:massbasis}. All momenta are flowing into the vertex.}
\end{figure}

The effect of HAC/EFT on the Higgs coupling to vector bosons is to introduce a non-trivial momentum dependence in the vertex, as one can see by inspecting the Feynman rules of the Higgs with $WW$, $ZZ$ and $Z\gamma$ vector bosons, which are presented in Figure~\ref{f:feynrules}. The tree-level SM piece is independent of the momentum, whereas New Physics potentially introduces new momentum-dependent Lorentz structures. New Physics can thus affect both rates of production and decay, as well as differential distributions. Exploiting differences in shape due to these new effects is one of the main avenues to look for indirect signals of New Physics. Here understanding SM higher-order corrections is especially important. This is primarily due to changes in shape arising from higher order matrix elements which are particularly relevant in the high-$p_T$ region.

\newcommand{\nr}{\stepcounter{diagram}(FR.\arabic{diagram})}
\newcounter{diagram}

\begin{fmffile}{feyngraph}
\end{fmffile}

We observe that in possible models which may generate these anomalous couplings, i.e. {\it UV completions}, not all Lorentz structures may be simultaneously generated. Indeed, in a large class of models, HAC of the type $g^{(2)}_{hvv}$ do not occur, e.g. in 2HDMs~\cite{withJosemi}, radion/dilaton exchange~\cite{withJosemi} or supersymmetric loops involving  sfermions or gauginos~\cite{withEduard}. 
This makes the study of phenomenology in which $g^{(2)}_{hvv}=0$ and $g^{(2)}_{hvv}\ne0$ particularly interesting. 

\begin{table}[t!]
\begin{center}
\begin{tabular}
{|c | c |c |}
\hline
 Operator & Coefficient & Constraints \\
\hline
${\mathcal O}_W=\frac{ig}{2}\left( H^\dagger  \sigma^a \lra {D^\mu} H \right )D^\nu  W_{\mu \nu}^a$ & $\frac{m_W^2}{\Lambda^2}(\bar c_W - \bar c_B)$ & (-0.035,0.005) \\
${\mathcal O}_B=\frac{ig'}{2}\left( H^\dagger  \lra {D^\mu} H \right )\partial^\nu  B_{\mu \nu}$ & $\frac{m_W^2}{\Lambda^2}(\bar c_W+\bar c_B)$  &  $(-0.0033,0.0018)$ \\
\hline
${\mathcal O}_{HW}=i g(D^\mu H)^\dagger\sigma^a(D^\nu H)W^a_{\mu\nu}$ & $\frac{m_W^2}{\Lambda^2}\bar c_{HW}$  & $(-0.035,0.015)$  \\
\hline
${\mathcal O}_{HB}=i g^\prime(D^\mu H)^\dagger(D^\nu H)B_{\mu\nu}$ & $\frac{m_W^2}{\Lambda^2}\bar c_{HB}$  & $(-0.045,0.075)$ \\
\hline
\end{tabular}
\end{center}
\caption{ Current $95\%$ CL Run1 constraints on the CP-even operators marginalized in a simultaneous fit~\cite{withJohn}.}
\label{tab:LHCoperators}
\end{table}

Finally, we comment on the current bounds for these operators from a global analysis including LEP and LHC Run 1 performed in Ref.~\cite{withJohn}, see Refs.~\cite{otherfits} for other examples of fits in this context. This analysis took into account all the CP-even operators including pure gauge and operators involving fermions, but not the CP-odd couplings $\tilde g$. In Table~\ref{tab:LHCoperators} we show the result of a global fit.
 When  $g^{(2)}_{hvv}=0$, i.e. $\bar c_W = - \bar c_{HW}$, the limits from a global fit of Run1 data are more stringent~\cite{withJohn} 
 \bea
 \bar c_{HW} = - \bar c_W= (0.004, 0.02) \ .
 \eea

\section{Calculation} \label{sec:Calc}

In this section, we present the calculation of the NLO QCD predictions for a Higgs produced in association with a vector boson in the presence of anomalous couplings. 
We first present the calculation of the fixed order results needed for the NLO implementation in MCFM, and then proceed to discuss the 
implementation in the {\sc Powheg-Box}. Our results are computed in terms of the modified Feynman rules presented in Figure~\ref{f:feynrules} such that the anomalous couplings are a function of $g^{(i)}_{hVV}$. 
At NLO accuracy the production and decay 
for $pp\rightarrow VH \rightarrow $leptons $+b\overline{b}$ completely factorize due to $SU(3)$ color structure. This is because gluon radiation linking the initial state to the final state has no contribution at NLO since its interference with the LO amplitude results in a contribution proportional to Tr($T^a)$, where $T_a$ is an $SU(3)$ generator. We therefore present amplitudes for $pp\rightarrow VH \rightarrow $leptons $H$, and allow the subsequent MC code to decay the Higgs boson (PYTHIA, or MCFM). In this paper the MCFM prediction corresponds to a LO decay, whilst the PYTHIA prediction includes effects from the parton shower. 

\subsection{Amplitudes for $WH$ production}

The LO amplitude for the associated production of a $W$ and Higgs boson has the following form, 
\begin{eqnarray} 
A^{(0)}_5(1_{q},2_{\overline{q}},3_{\ell},4_{\overline{\nu}},H) = \left(\frac{g}{\sqrt{2}}\right)^2\delta^{i_1}_{j_2}\mathcal{A}^{(0)}_5(1_{q},2_{\overline{q}},3_{\ell},4_{\overline{\nu}},H),
\label{eq:WHLOAmp}
\end{eqnarray}
where we have defined the full amplitude $A^{(0)}_5$ in terms of a color stripped primitive amplitude $\mathcal{A}^{(0)}_5$; at LO the color factor is simply $\delta^{i_1}_{j_2}$. 
In addition to the extraction of the overall color factor in eq.~\ref{eq:WHLOAmp} we have also extracted the electroweak pre-factors from the fermionic $W$ vertices. We 
note however that we have not extracted the $g$ from the $HWW$ vertex since this will be modified during our calculation. Although we will consider decays of the Higgs to $b\overline{b}$, for simplicity we suppress the decay of the Higgs in this section. 

We will also require the amplitudes needed to construct the NLO corrections. This consists of two new amplitudes, the one-loop virtual amplitude $A^{(1)}_5$
and the real emission amplitude containing an additional parton  $A^{(0)}_6$(in this case an additional gluon). The virtual primitive amplitude is defined as follows, 
\begin{eqnarray} 
A^{(1)}_5(1_{q},2_{\overline{q}},3_{\ell},4_{\overline{\nu}},H) = \left(\frac{g}{\sqrt{2}}\right)^2\delta^{i_1}_{j_2}\;\frac{\alpha_s}{4\pi}\left(\frac{N_c^2-1}{N_c}\right)\mathcal{A}^{(1)}_5(1_{q},2_{\overline{q}},3_{\ell},4_{\overline{\nu}},H) 
\label{eq:WHVAmp}
\end{eqnarray}
The real emission amplitude including the parton $7_g$\footnote{Our naming convention follows the implementation in MCFM such that $p_5$ and $p_6$ are reserved for the decay of the Higgs boson to $b\overline{b}.$} is thus 
\begin{eqnarray}
A^{(0)}_6(1_{q},2_{\overline{q}},3_{\ell},4_{\overline{\nu}},H,7_g) =g_s \left(\frac{g}{\sqrt{2}}\right)^2(T^{a_7})^{i_1}_{j_2}\;\mathcal{A}^{(0)}_6(1_{q},2_{\overline{q}},3_{\ell},4_{\overline{\nu}},H,7_g) 
\label{eq:WHVAmp}
\end{eqnarray}
The color stripping for the real emission amplitude is slightly more complicated than the LO and depends on the color matrix $T^{a_7}$, however there is still only one unique 
color ordering which simplifies the calculation significantly. 

Since we are interested in associated production we are able to factorize the QCD 
corrections which affect the initial state, from the modifications to the $HVV$ vertex. 
The factorization proceeds as follows for all of the primitive amplitudes we have considered,   
\begin{eqnarray}
\mathcal{A}^{(i)}_j ( p_1+p_2 \rightarrow V_1(\rightarrow H + V_2) + X) = \mathcal{J}^{\mu}_{SM}(p_1,p_2,V_1,X){P}^{V_1}_{\mu\nu}(P_{12X})V^{\nu}_{\Lambda}(V_2,H) 
\label{eq:curfac}
\end{eqnarray}
In the above equation $\mathcal{J}^{\mu}_{SM}(p_1,p_2,V_1,X)$ represents the production of a (chiral) current in the SM, if $j=5$ then $X=0$, whilst for the 
real emission amplitude $j=6$ and $X$ corresponds to the emission of an additional gluon. The second current $V^{\nu}_{\Lambda}(V_2,H)$ corresponds to the splitting of the initial vector boson $V_1$ into $V_2$ and $H$, with the subsequent decays of $V_2$ to leptons included.
Finally the two currents are connected by vector boson propagator, which in the unitary gauge is defined as follows, 
\begin{eqnarray}  
 P^{W}_{\mu\nu}(k) = -\frac{i}{k^2-m_W^2}\left(g_{\mu\nu} - \frac{k_{\mu}k_{\nu}}{M_W^2}\right)
\end{eqnarray}
Since the $W$ bosons decay we work in the complex mass scheme. 
However, in this calculation the longitudinal $k_{\mu}k_{\nu}$ pieces do not contribute since $\mathcal{J}^{\mu}_{SM}(p_1,p_2,V_1,X)P^{12X}_{\mu} = 0$ for massless initial and final states.
We also frequently use the following function in our calculations  
\begin{eqnarray}
\mathcal{P}_V(s) = \frac{s}{s-M_V^2+i M_V \Gamma_V}
\end{eqnarray}


In order to complete our discussion of the amplitudes for $WH$ production we present results for the currents defined above. 
Since we are discussing $W$ bosons, it is natural to relate the chiral currents to helicity amplitudes. Therefore we will use the language and notation 
of the spinor helicity formalism in the following definitions. 
We refer readers unfamiliar with the spinor-helicity formalism to one of the many reviews. For instance a detailed introduction can be found in ref.~\cite{Dixon:1996wi}.

The first current we define is the modified decay current including the effect of the dimension-6 operators $\mathcal{W}^{\mu}_{\Lambda}(P_{2},3^{-}_{\nu},4^{+}_{e^+},H_{234})$, where $P_2$ is the inflowing momenta, $H_{234}$ is the outgoing Higgs boson, and $p_3$ and $p_4$ represent the momenta of the final state leptons. 
The explicit form for this current is as follows, 
\begin{eqnarray}
\mathcal{W}^{\mu}_{\Lambda}(P_{2},3^{-}_{\nu},4^{+}_{e^+},H_{234})&=& \frac{i\mathcal{P}_W(s_{34})}{s_{34}}\bigg\{
\spab 3.\gamma^{\mu}.4 (g m_W + g^{(2)}_{hWW}( P_2^2+s_{34})) - g^{(2)}_{hWW} P_2^{\mu} \spab 3.P_2.4
\nonumber\\ &&-\frac{(g^{(1)}_{hWW}-i\tilde{g}_{hWW})}{2}
\left(\spaa 3.\gamma^{\mu}.P_2.3\spb4.3+\spa3.4\spbb 4.P_2.\gamma^{\mu}.4\right)
\bigg\}
\label{eq:oneoffshell}
\end{eqnarray}
Next we consider the chiral currents in the SM. The current needed for the construction of the LO amplitude is rather simple,  
\begin{eqnarray}
\mathcal{J}^{\mu}_{LO}(1^-_{u},2_{\overline{d}}^+,P_{12}) = -i \spab 1.\gamma^{\mu}.2
\end{eqnarray} 
The virtual current corresponding to a vertex correction, is also very simple, 
\begin{eqnarray}
\mathcal{J}^{\mu}_{Virt}(1^-_{u},2_{\overline{d}}^+,P_{12}) &=&  2\left(
-\frac{1}{\epsilon^2}\left(\frac{\mu^2}{-s}\right)^{\epsilon}-\frac{3}{2\epsilon}\left(\frac{\mu^2}{-s}\right)^{\epsilon}+\frac{\pi^2}{2}-\frac{7}{2}\right) \nonumber\\ && \times \mathcal{J}^{\mu}_{LO}(1^-_{u},2_{\overline{d}}^+,P_{12})
\end{eqnarray} 
Finally the current corresponding to the emission of an additional gluon, necessary in the real calculation has two possible helicity configurations, 
\begin{eqnarray}
\mathcal{J}^{\mu}_{Real}(1^-_{u},2_{\overline{d}}^+,3_g^+,P_{123})=- i  \frac{\spaa 1.P_{123}.\gamma^{\mu}.1}{\spa2.3\spa3.1}  \\
\mathcal{J}^{\mu}_{Real}(1^-_{u},2_{\overline{d}}^+,3_g^-,P_{123})=- i \frac{\spbb 2.P_{123}.\gamma^{\mu}.2}{\spb2.3\spb3.1}  
\end{eqnarray}
Contracting these various currents together results in the amplitudes for the production of $WH$ at NLO in QCD including the effects of the dimension-6 operators. For example the LO contraction is, 
\begin{eqnarray} 
 A^{(0)}_{5}(1^-_{u},2_{\overline{d}}^+,3^-_{\nu},4_{e^+}^+,H_{1234})=\mathcal{J}^{\mu}_{LO}(1^-_{u},2_{\overline{d}}^+,P_{12}) P^{W}_{\mu\nu}(P_{12})\mathcal{W}^{\nu}_{\Lambda}(P_{12},3^{-}_{\nu},4^{+}_{e^+},H_{1234})
 \end{eqnarray}

\subsection{Amplitudes for $ZH$ production}

Next we consider the production of a $Z$ boson in association with a Higgs boson. The situation is slightly more complicated than 
the $WH$ example considered in the previous section as the internal boson can be either a $Z$ or a virtual photon $\gamma^*$. 
In the SM the latter case does not occur, but the full general anomalous coupling parametrization allows for this contribution. We therefore parametrize 
the LO amplitude as follows, 
\begin{eqnarray} 
A^{(0)}_5(1_{q},2_{\overline{q}},3_{\ell},4_{\overline{\ell}},H) = 2 e^2 \delta^{i_1}_{j_2}\sum_{i,j = L,R} v^\ell_j \bigg(Q_q \mathcal{A}^{(0)\gamma}_5(1_{q},2_{\overline{q}},3_{\ell},4_{\overline{\ell}},H)_{ij} \nonumber\\+ v^q_j \mathcal{A}^{(0)Z}_5(1_{q},2_{\overline{q}},3_{\ell},4_{\overline{\ell}},H)_{ij}\bigg)
\label{eq:ZHLOAmp}
\end{eqnarray} 
Unlike the case for $WH$ there are now four allowed helicity configurations, corresponding to the selection of $L$ and $R$ helicities for particles 1 and 3. 
The left and right handed couplings (in the SM) are defined as follows, 
\begin{eqnarray} 
v_{L}^{\ell} = \frac{-1-2Q_{\ell}\sin^2\theta_W}{\sin 2\theta_W}, \quad v_R^\ell = - \frac{2Q_{\ell}\sin^2\theta_W}{\sin 2\theta_W} \\
v_{L}^{q} = \frac{\pm 1-2Q_{q}\sin^2\theta_W}{\sin 2\theta_W}, \quad v_R^q = - \frac{2Q_{q}\sin^2\theta_W}{\sin 2\theta_W}. 
\end{eqnarray} 
The sign in the $v_L^q$ distinguishes between up (+) and down (-) type quarks. 
The amplitudes involving $Z$ exchange can be obtained from the results presented  in the previous section. The results for the analogous 
case in which $W\rightarrow Z$ can be obtained by simply swapping $g\rightarrow \frac{g}{\cos\theta_W^2}$ and $g^{(i)}_{hWW} \rightarrow g^{(i)}_{hZZ}$ in eq.~\ref{eq:oneoffshell}. The results then correspond to the $LL$ configuration, other configurations can be obtained from fermion line reversal symmetries. 
The current for a virtual photon exchange is given by, 
\begin{eqnarray} 
\mathcal{J}^{\mu}_{\gamma^*}(P_{2},3^{-}_{\ell},4^{+}_{\overline{\ell}},H_{234})&=& \frac{i\mathcal{P}_Z(s_{34})}{s_{34}}\bigg\{
\frac{(g^{(1)}_{haZ}-i\tilde{g}_{haZ})}{2}(\spab 3.\gamma^{\mu}.4(\spab 3.2.3 + \spab 4.2.4) \nonumber\\ && - 2 (p_3^{\mu}+p_4^{\mu})\spab 3.2.4)  + 
g^{(2)}_{haZ}\left(\spab 3.\gamma^{\mu}.4 p_2^2- (p_2^{\mu})\spab 3.2.4\right) 
\bigg\}
\end{eqnarray}
The SM currents are related to those described in the previous section.  

\subsection{Implementation in Monte Carlo codes\label{sec:implementation}}

The amplitudes calculated above have been implemented into the parton level Monte Carlo code MCFM. Infrared divergences are regulated using the Catani Seymour 
Dipole subtraction method~\cite{CataniSeymour}. We use the default MCFM electroweak parameters, which correspond to the following choices, 
\begin{eqnarray} 
M_Z = 91.1876 \; {\rm{GeV}}, \quad M_W = 80.398 \; {\rm{GeV}}, \\ 
\Gamma_Z = 2.4952 \; {\rm{GeV}}, \quad \Gamma_W = 2.1054 \; {\rm{GeV}}, \\ 
G_F = 0.116639 \times 10^{-4} \; {\rm{GeV}^{-2}}, \quad m_t =173.2 \;{\rm{GeV}} 
\end{eqnarray} 
The remaining EW parameters are defined in terms of the above input parameters. Since we make the choice of defining the input of our program in terms of the dimension-6 Wilson coefficients of eq.~\eqref{eq:SILH}, some additional effects are taken into account to fully map the physical effects of the EFT description into our HAC Lagrangian. Of the operators that we consider in our implementation -- $\mathcal{O}_{W},\mathcal{O}_{B}$ and $\mathcal{O}_{\gamma}$ -- lead to non-canonical $SU(2)\times U(1)$ gauge boson kinetic terms after electroweak symmetry breaking. The field and gauge coupling redefinitions necessary to canonically normalise the theory lead to $\mathcal{O}(\Lambda^{-2})$ modifications of both the EW parameters in terms of the inputs as well as the couplings of gauge bosons to fermions as compared to the usual SM expectations. There is some freedom in how these redefinitions are performed and therefore the places in which these corrections appear although physical observables are naturally independent of such choices at this order in the EFT expansion. Appendix~\ref{app:redefs} describes the choices we make and therefore the origin of the relations and corrections that follow.

For the EW parameters, we work in the $m_W, m_Z, G_F$ scheme, and define the SM values for the Weinberg angle, electric charge and Higgs v.e.v as
\begin{eqnarray}
\tilde{c}_W  = \frac{m_W}{m_Z}, \quad \tilde{e} = 2\frac{m_W}{v} \sqrt{1-\frac{m_W^2}{m_Z^2}}, \quad v^2 = \frac{1}{2G_F}.
\end{eqnarray}
These are corrected due to a relative shift in the $Z$-boson mass
\begin{eqnarray}
\delta m_Z = \frac{\tilde{e}^2}{8\tilde{c}^2_W} \frac{v^2}{m_W^2} \left(2 \overline{c}_B + \tilde{c}_W^2 \overline{c}_W \right),
\end{eqnarray} 
which redefines the mixing angle and $\alpha(m_Z)$ as follows, 
\begin{align}
\hat{c}_W^2 &= \tilde{c}_W^2 (1 + 2\delta m_Z) \\
e &= \tilde{e}\left(1-\frac{\tilde{c}^2_W}{\tilde{s}_W^2}\delta m_Z\right)=\sqrt{4\pi\alpha(m_Z)},
\end{align}
While the definition of the Higgs v.e.v in terms of $G_F$ is unchanged.
All other EW parameters are derived from the modified values $\hat{c}_W^2$ and $e$ using SM relations.

The coupling between a photon and fermion is corrected by the term, 
\begin{eqnarray}\label{eq:ashift}
\delta e = - \frac{v^2}{m_W^2}\frac{e^2}{8} \overline{c}_W
\end{eqnarray}
while the left and right handed couplings of the $Z$ to a fermion, $f$, with weak isospin $T_3^f$ and electric charge $Q^f$ are shifted as follows,  
\begin{eqnarray}\label{eq:zshift}
\delta f_L^Z &=& \frac{\tilde{e}}{\tilde{c}_W\tilde{s}_W} \left(T_3^f \delta T_3^Z - Q^f \tilde{s}_W^2 \delta Q^Z\right), \\
\delta f_R^Z &=& -\tilde{e}\frac{\tilde{s}_W}{\tilde{c}_W} Q^f \delta Q^Z,
\end{eqnarray}
with
\begin{eqnarray}
\delta T_3^Z = \delta m_Z,\quad
\delta Q^Z = \frac{v^2}{m_W^2} \frac{e^2}{8 \tilde{s}^2_W \tilde{c}^2_W} (2 \overline{c}_B + \tilde{c}^4_W \overline{c}_W).
\end{eqnarray}
Phenomenologically speaking, the effect of these re-definitions is minor, and typically results in changes which are sub-precent 
compared to predictions which do not alter the EW parameters by the EFT operators. For the choice of parameters simulated in this paper, only $\overline{c}_W$ affects the EW parameters and neutral gauge boson couplings. 

The Higgs width is also modified as a result of the anomalous interactions, and
we use the {\sc eHDECAY} implementation described in ref.~\cite{continos} to define the modifications to the Higgs width. 

The {\sc Powheg-Box}~\cite{Frixione:2007vw,Alioli:2010xd} provides a mechanism to match fixed order results to parton shower level Monte Carlo codes.  In our case the implementation 
is rather straightforward, in particular since associated production in the SM has already been considered in the literature~\cite{Hamilton:2009za,Luisoni:2013kna} . Therefore to include 
our results in the {\sc Powheg-Box} we have simply updated the existing matrix elements with our own calculations described earlier in this section. The only 
technical task is to ensure that all of the variables required in the MCFM matrix element routines, are appropriately initialized by to the values assigned in {\sc Powheg}. 
This is achieved through an interface to the MCFM routines which is called once at runtime.

\section{Results}\label{sec:results}

\subsection{Fixed order results\label{sec:FO}} 

In this section we present results obtained at fixed order in perturbation theory. Specifically we study the dependence on the total 
rate at NLO as a function of the anomalous couplings. In order to simplify our results we do not vary all parameters continuously. Instead
we focus on parameters which are representative of the phenomenology at the LHC. Ignoring for now the CP-violating operators we note 
that the in our basis the variables $\overline{c}_W$ and $\overline{c}_{HW}$ are sufficient to probe the Lorentz structures of the Feynman rule
associated with $g^{(1)}_{hww}$ and $g^{(2)}_{hww}$. In particular $\overline{c}_{HW} = 0$ and $\overline{c}_W \ne 0$ probes the regime 
in which $g^{(2)}_{hww}$ modifies the vertex, and if $\overline{c}_W=-\overline{c}_{HW}$ then the regime in which $g^{(1)}_{hww}$ modifies the vertex 
is selected. 

In order to study the impact of the anomalous couplings we calculate cross sections for the LHC at the 13 TeV in which the final state particles have to satisfy the following phase 
space selection criteria, 
\begin{eqnarray}
WH  &:& \nonumber\\
&& p_T^{\ell} > 25 \; {\rm{GeV}}, \quad |\eta_{\ell} | < 2.5, \quad {\rm{MET}} > 45 \; {\rm{GeV}}, \quad 2 \; b\,{\rm{jets}} : \; p_T^j > 25 \; {\rm{GeV}}, \quad |\eta_b| <2.5  \nonumber\\
ZH  &:& \nonumber\\
&& p_T^{\ell} > 25 \; {\rm{GeV}} \quad |\eta_{\ell} | < 2.5    , \quad 2 \; b\,{\rm{jets}} : \; p_T^j > 25 \; {\rm{GeV}}, \quad |\eta_b| <2.5. \nonumber
\end{eqnarray}
We refer to this selection as our basic-cuts. Since the effects of the EFT operators are more apparent at higher energies we also define a high-$p_T^V$ selection cut 
in which we impose an additional cut on $p_T^V > 200$ GeV. We note that $p_T^V$ is a well defined experimental observable for both $W$ and $Z$ processes. 

\begin{figure} 
\begin{center} 
\includegraphics[width=7cm]{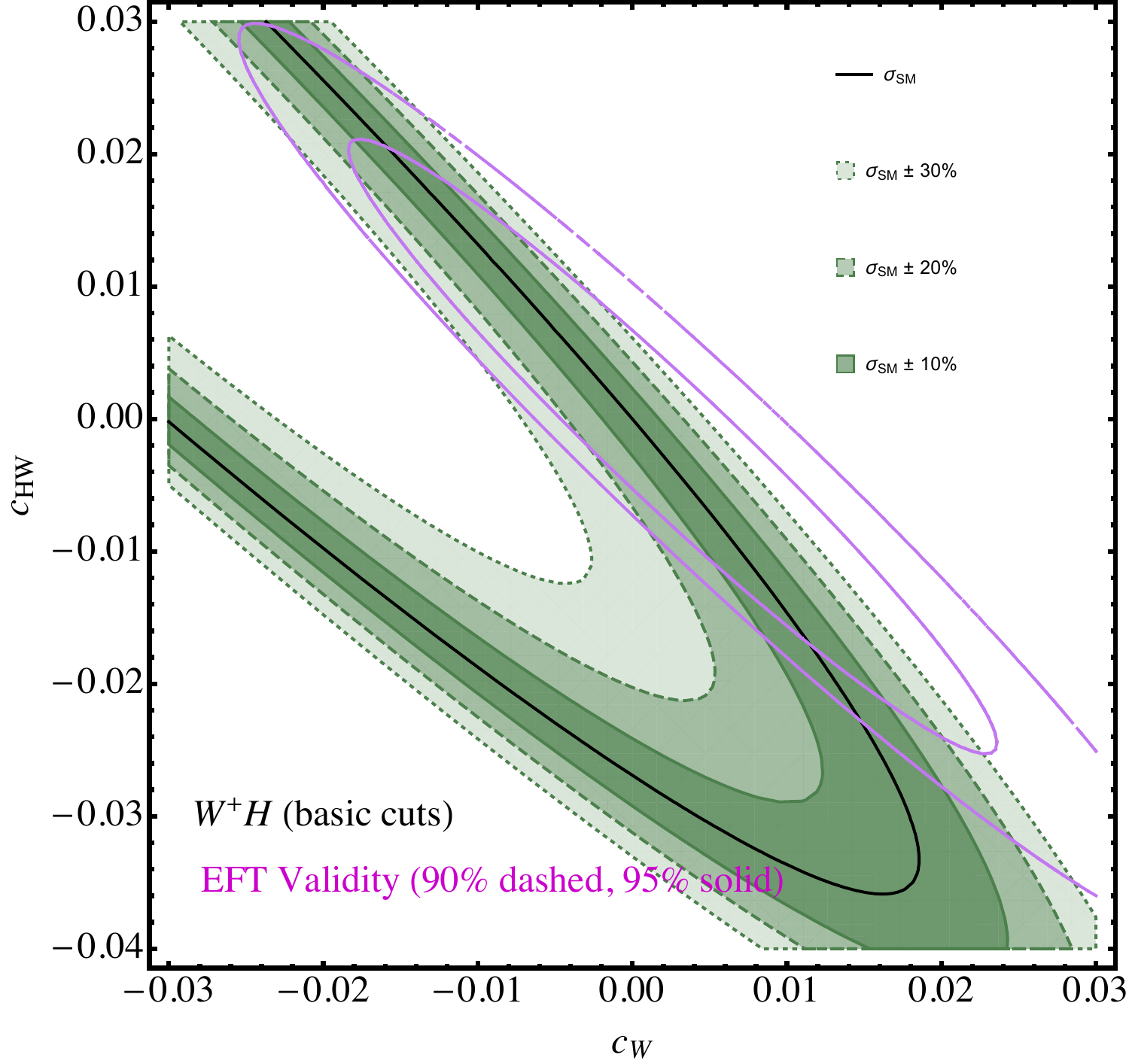} \includegraphics[width=7cm]{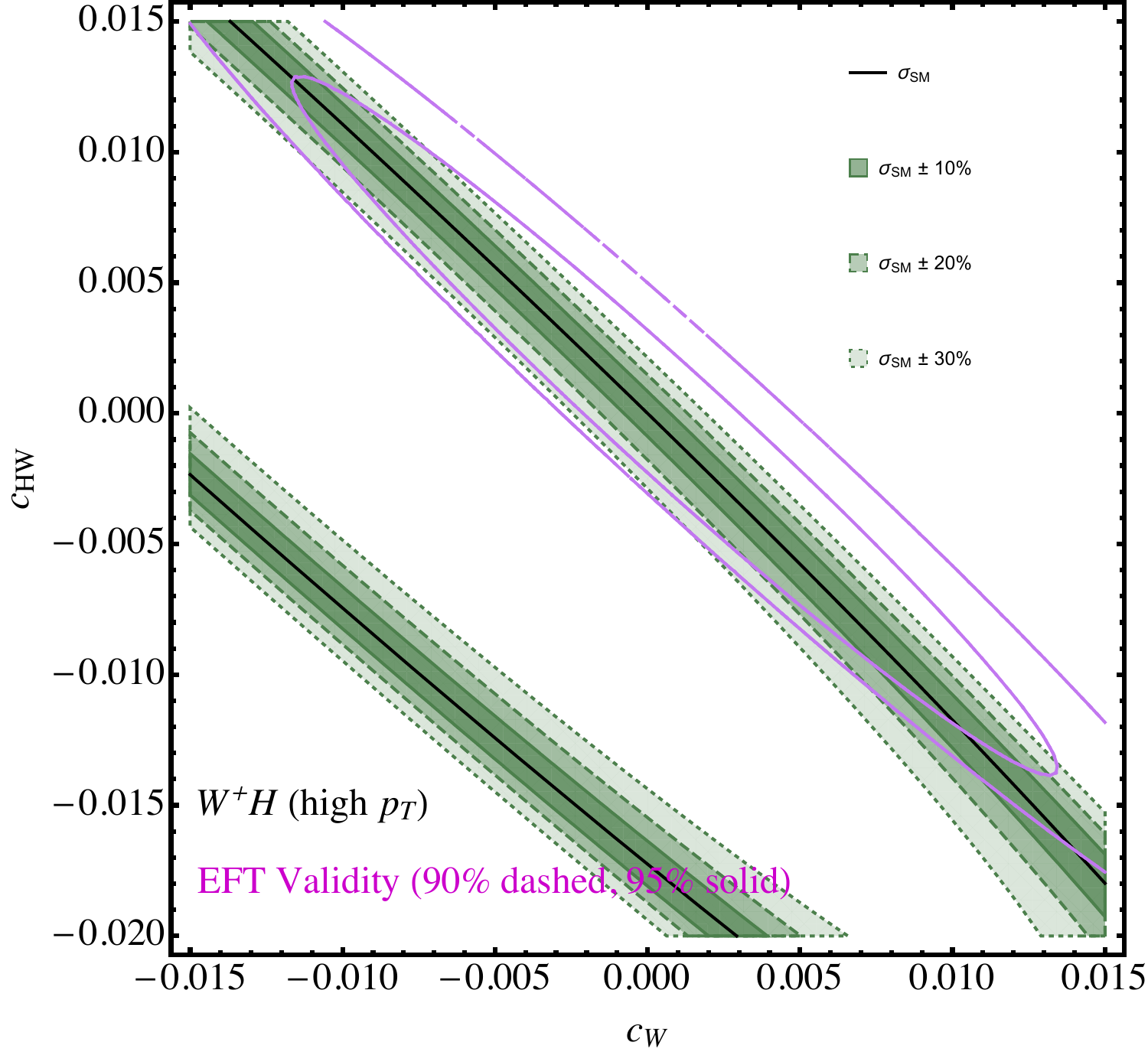} 
\caption{Contours representing deviations as a function of $\overline{c}_W$ and $\overline{c}_{HW}$ from the NLO SM prediction  for $W^+H$  production at the 13 TeV LHC. The plots on the left hand side correspond to the basic cut selection, whilst those on the right include an additional cut on $p_T^Z > 200$ as described in the text.}
\label{fig:WHcwchw}
\end{center} 
\end{figure}

\begin{figure} 
\begin{center} 
\includegraphics[width=7cm]{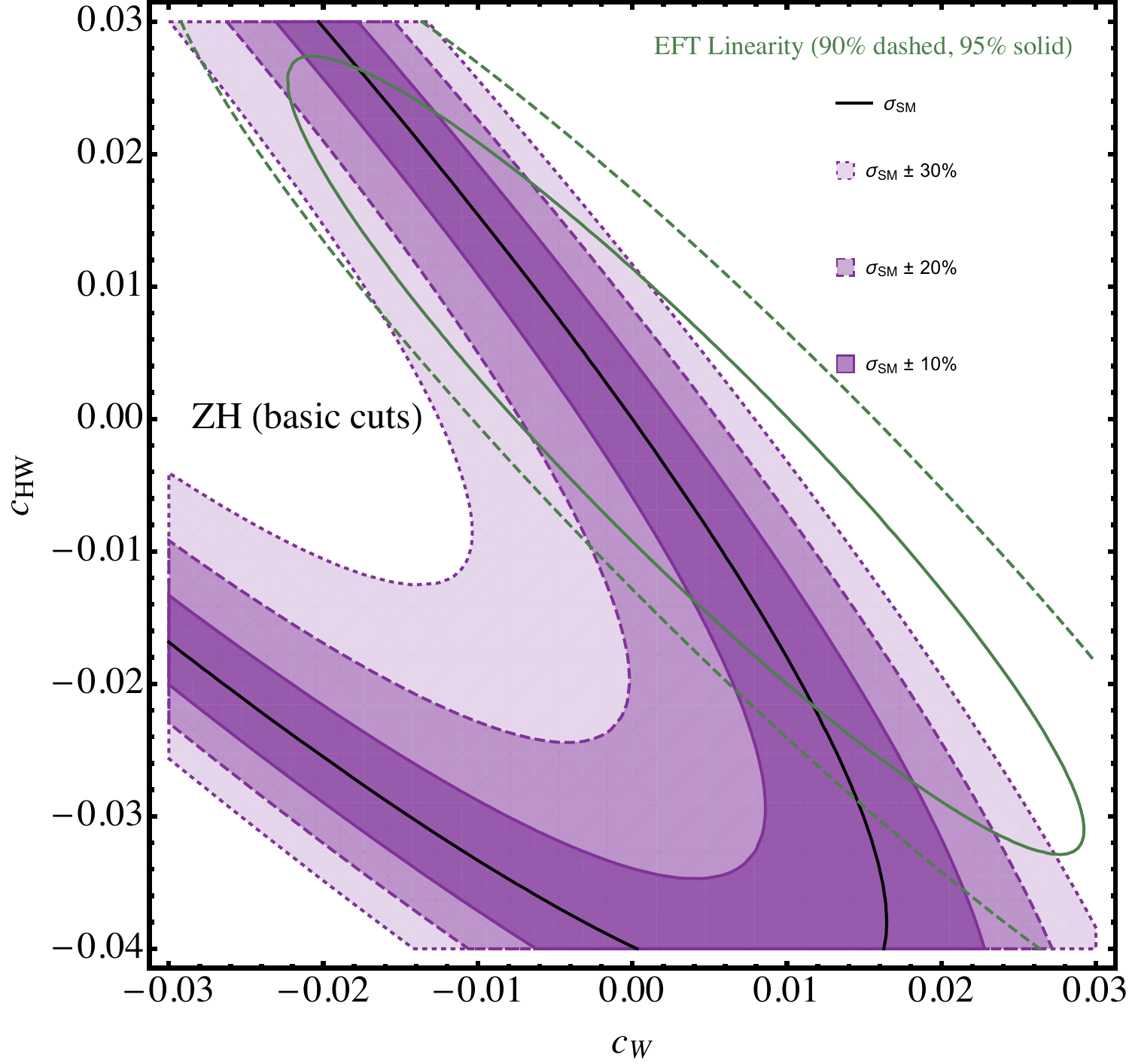} \includegraphics[width=7cm]{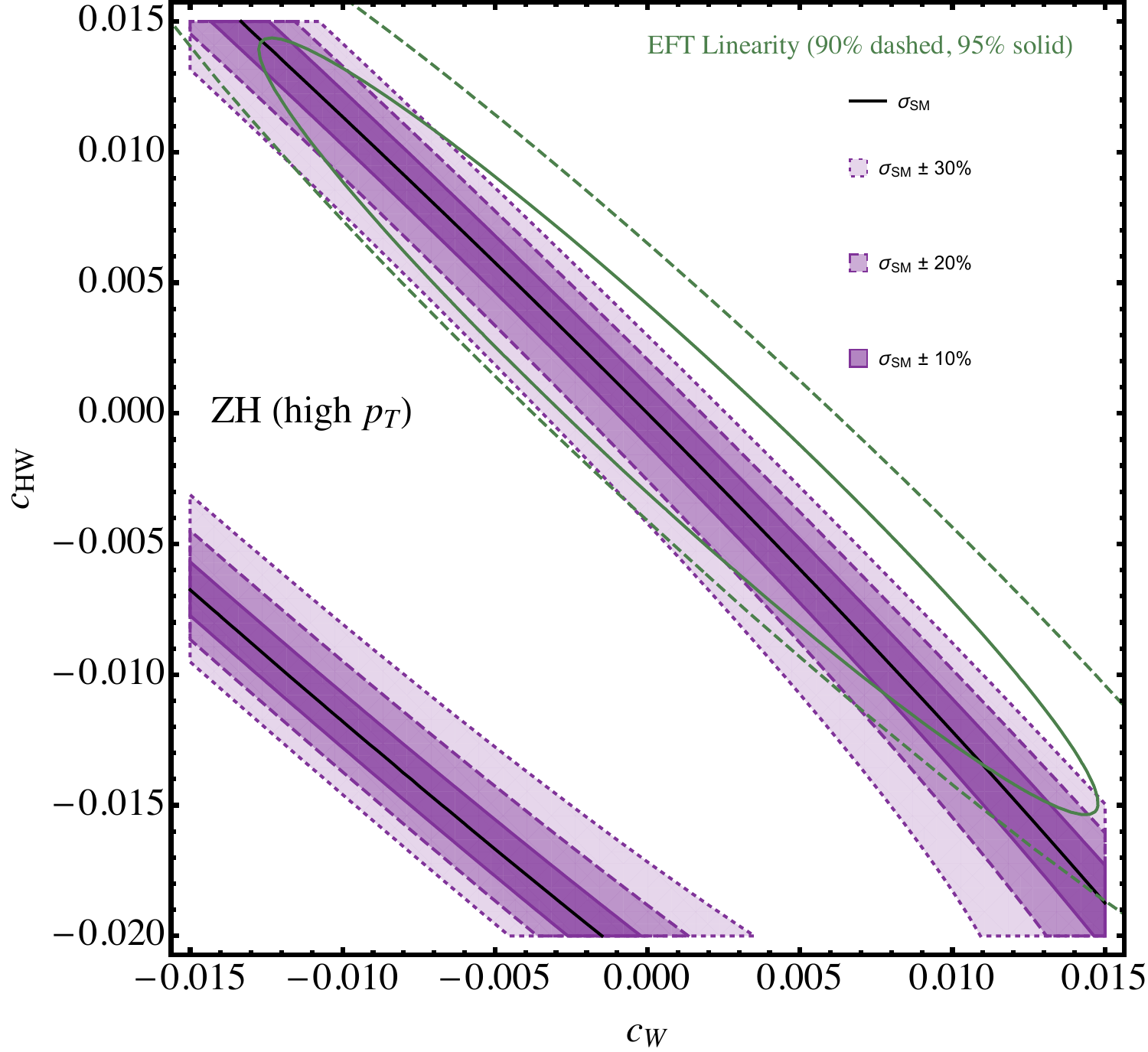} 
\caption{Contours representing deviations as a function of $\overline{c}_W$ and $\overline{c}_{HW}$ from the NLO SM prediction for $ZH$  production at the 13 TeV LHC. The plots on the left hand side correspond to the basic cut selection, whilst those on the right include an additional cut on $p_T^Z > 200$ as described in the text. }
\label{fig:ZHcwchw}
\end{center} 
\end{figure}

Our results for the basic and high-$p_T^V$ cuts are presented in Figures~\ref{fig:WHcwchw} and~\ref{fig:ZHcwchw} for $WH$ and $ZH$ processes respectively. The results have been obtained using MCFM, with the parameters described in the previous section. We use the CT10 PDF set~\cite{Lai:2010vv} for NLO predictions and CTEQ6L1 for LO predictions. The renormalization and factorization scale have been set to $\mu=m_{VH}$.  In each of Figures~\ref{fig:WHcwchw} and~\ref{fig:ZHcwchw} the plot on the left side 
represents the cross section obtained using the basic cuts, whilst those on the right hand side correspond to the cross sections obtained with the additional high-$p_T^V$ selection requirement applied. In both figures we plot the cross section as a function of $\overline{c}_W$ and $\overline{c}_{HW}$. We present contours which 
correspond to the values of $\overline{c}_W$ and $\overline{c}_{HW}$ needed to obtain a $10$, $20$ or $30$ \% deviation from the SM prediction. For reference using our cuts described above the SM predictions are : $9.7$ fb and 1.8 fb for $W^+H$ with the basic and high-$p_T^V$ cuts respectively, and $5.1$ and $0.54$ for $ZH$.
 Our results contain terms up to order $\overline{c}_{X}^2$  as can be clearly seen from the figures, since the results for constant cross section form ellipses. Including these terms is somewhat dangerous, since in general they correspond to regions in which the EFT may be breaking down. This is because 8 dimensional operators also contribute first 
at this level, and therefore should not be ignored in the calculation. Therefore in an attempt to quantify the range of validity of our results we present a contour which corresponds to the regime in which the linear part of the cross section corresponds to 95\% (solid) and $90\%$ (dashed) of the total, or in other words the higher order pieces in the EFT should be small (both from 8 dimensional operators and loop corrections to the 6 dimensional operators which go like quadratic pieces).  Experimental results can then be used to set reliable EFT bounds inside of this contour. We stress that values which lie outside of this contour (i.e. larger absolute values of $\overline{c}_X$) cannot be reliably excluded given our theoretical accuracy, and given the form of our results, it is clear that there are regions outside of the EFT validity which conspire to produce small changes in the total cross section. 

The hallmark of EFT operators is a lack of suppression at high energies due to poor high-energy behaviour. Therefore, a natural place to search for the impact of 
the higher dimensional operators is the region which is sensitive to larger values of $\hat{s}$. Since $m_{VH}$ cannot be directly cut upon in the experiment for leptonic 
$WH$ final states, we use $p_T^V$ as a proxy for $\hat{s}$. The plots on the right hand side of Figures~\ref{fig:WHcwchw} and~\ref{fig:ZHcwchw} present these results. 
As expected we see a significant increase in sensitivity in the ($\overline{c}_{HW},\overline{c}_{W}$) plane compared to the more inclusive analysis. By looking at high $p_T^V$ one improves the sensitivity from around $|\overline{c}_W + \overline{c}_{HW}| \lsim 0.005$ to around $|\overline{c}_W + \overline{c}_{HW}| \lsim 0.002$.

\begin{figure} 
\begin{center} 
\includegraphics[width=7cm]{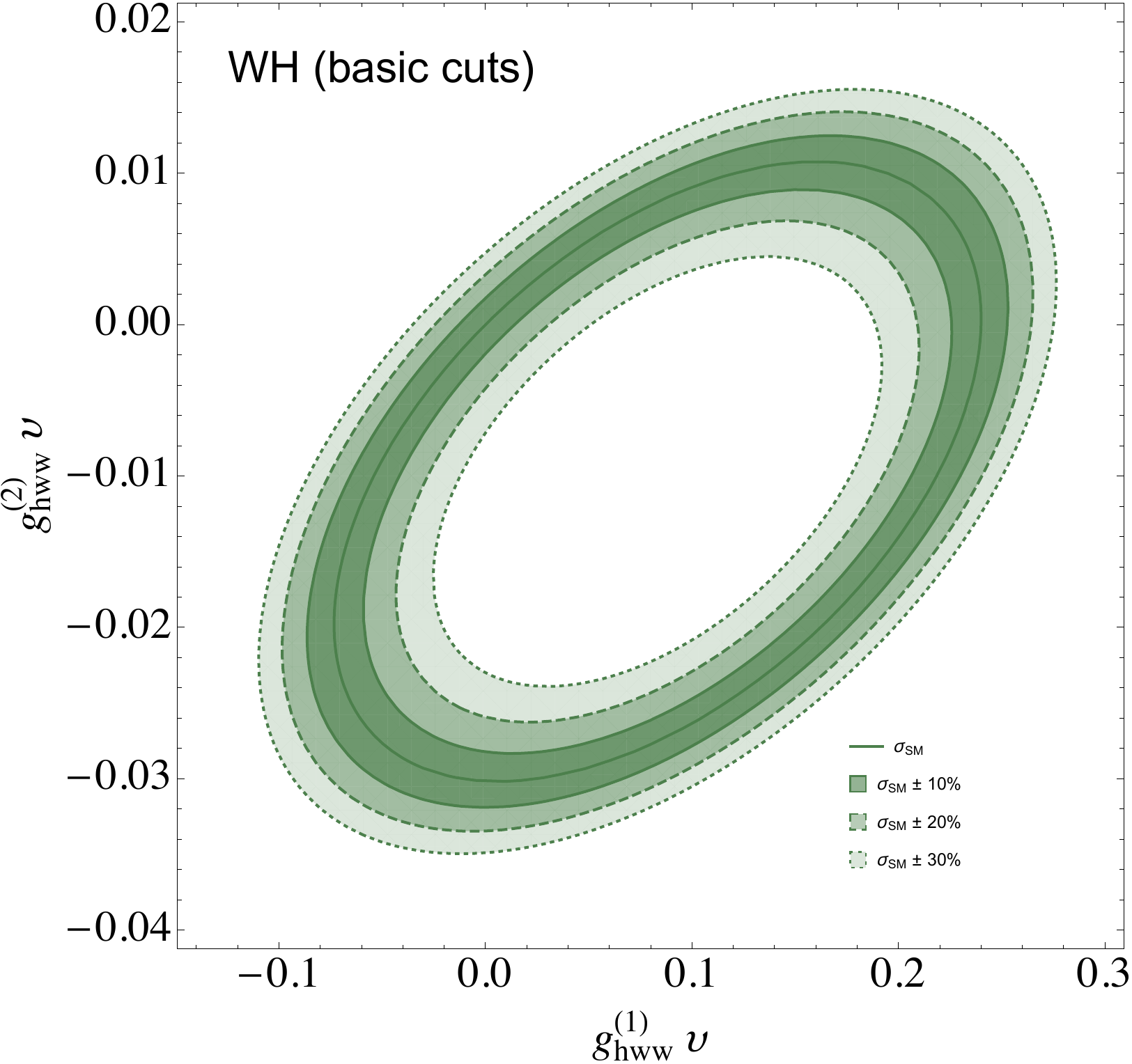} \includegraphics[width=7cm]{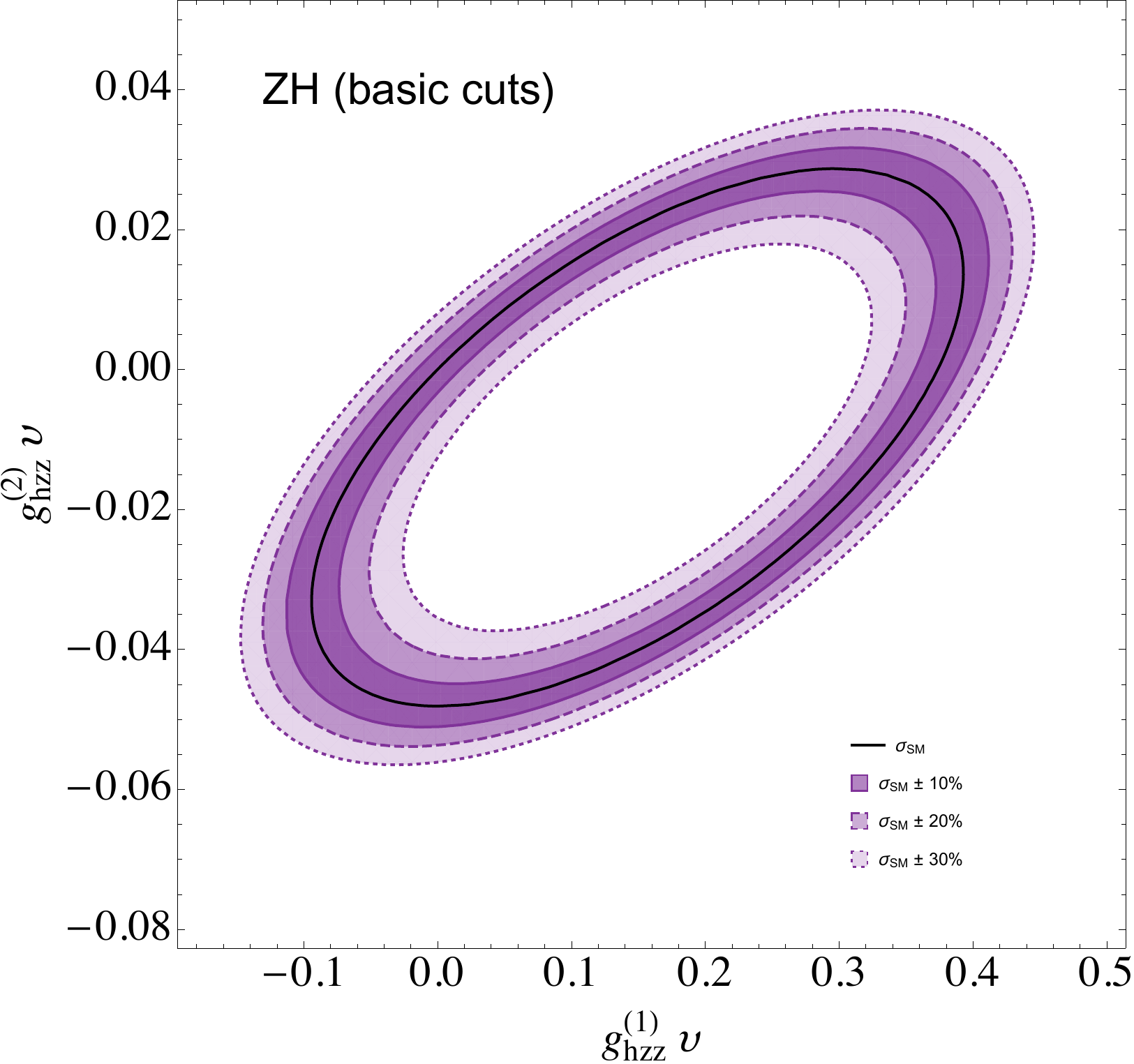} 
\caption{Contours representing deviations as a function of the Higgs anomalous couplings $g^{(1)}_{hvv} v$ and $g^{(2)}_{hvv} v$, where $v$ is the Higgs vev, from the NLO SM prediction for $WH$ (left) and $ZH$ (right)  production at the 13 TeV LHC. }
\label{fig:gplots}
\end{center} 
\end{figure}

To demonstrate the flexibility of our code,  we present results in the HAC basis, rather than the EFT approach, in Figure~\ref{fig:gplots}. In this setup the anomalous couplings are parameterized in terms of general Lorentz invariant operators in the Lagrangian. In this approach corrections from higher dimensional operators are not a concern so we present the full ellipses, and do not present EFT validity contours for these plots. We note that, since the HAC basis does not necessitate deviations in the EW parameters due to kinetic mixing of operators, we use the SM EW parameter choices (corresponding to the MCFM defaults) for these plots. 

\begin{figure} 
\begin{center} 
\includegraphics[width=7cm]{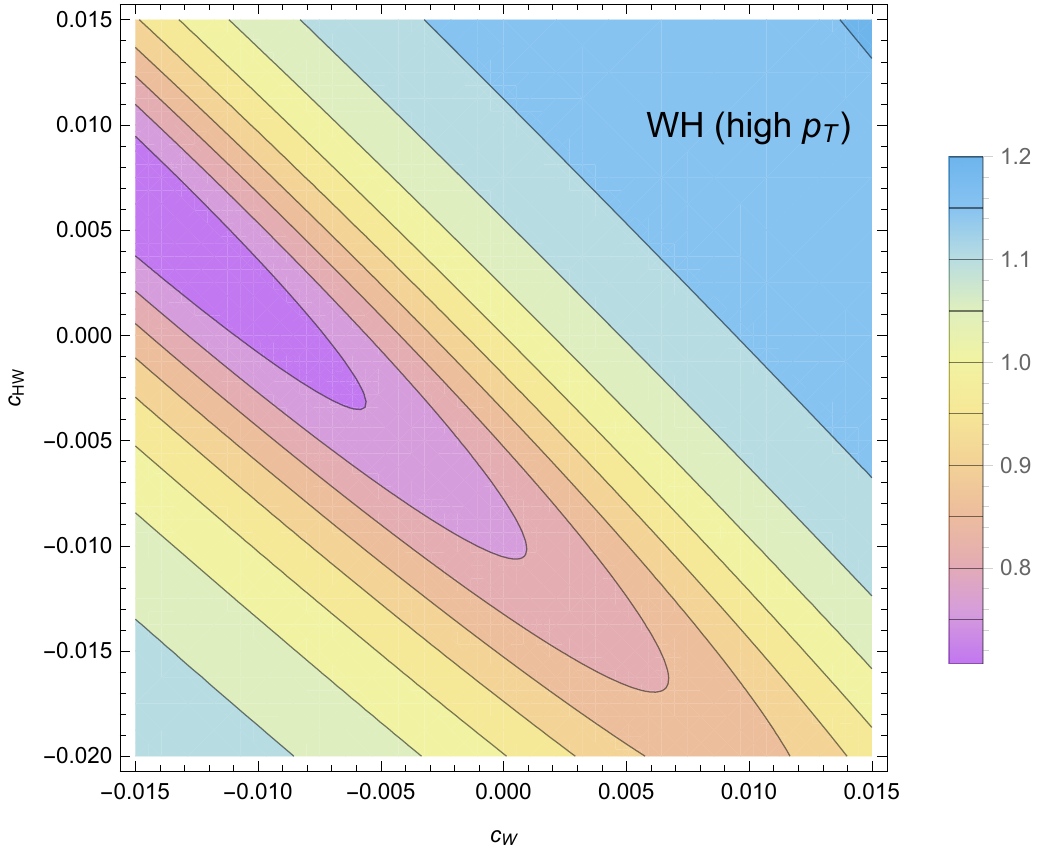} 
\includegraphics[width=7cm]{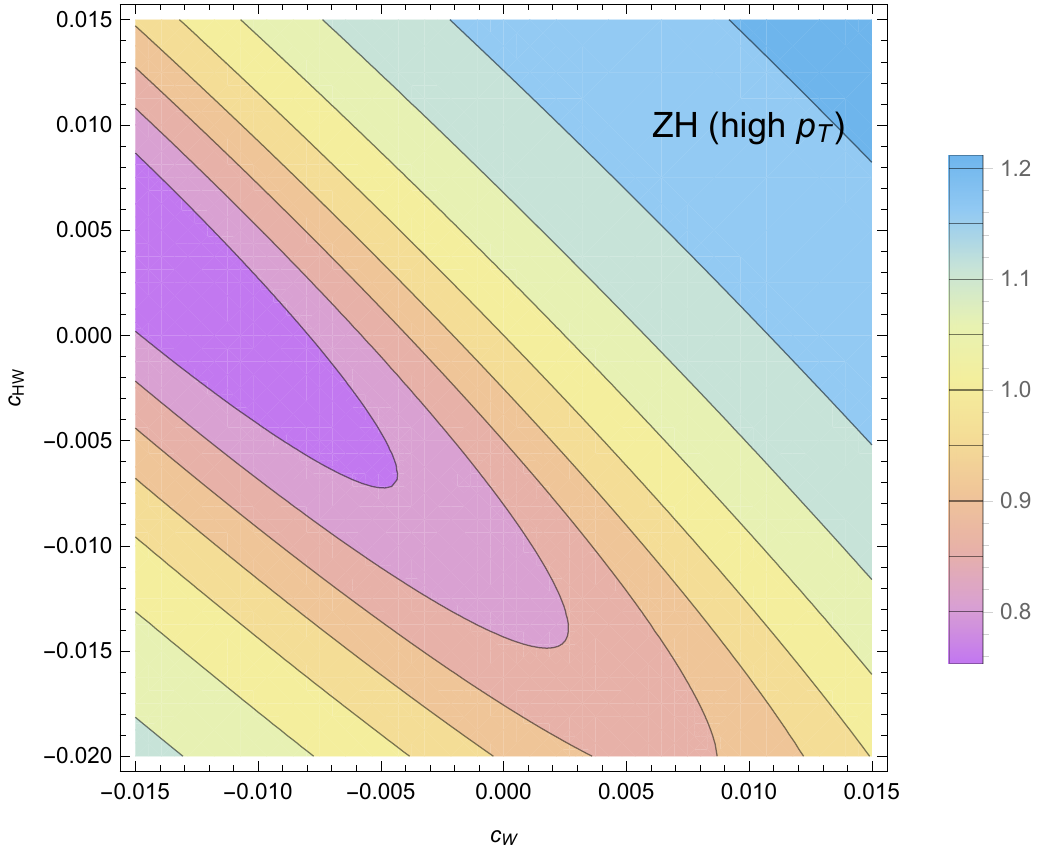} 
\caption{Relative impact of the NLO BSM corrections for $ZH$, high- $p_T^V$ selection cuts, i.e. $\mathcal{R}^{NLO}(\overline{c}_W,\overline{c}_{HW})$ as defined in the text, for both predictions the SM piece is included at NLO. }
\label{fig:bsmNLO}
\end{center} 
\end{figure}

In Figure~\ref{fig:bsmNLO} we investigate the impact of the NLO corrections to the anomalous couplings. We define the following ratio, 
\begin{eqnarray}
\mathcal{R}^{NLO}(\overline{c}_W,\overline{c}_{HW})=\frac{\sigma^{NLO}(\overline{c}_W,\overline{c}_{HW})}{\sigma^{NLO}(0,0)+\sigma^{LO}(\overline{c}_W,\overline{c}_{HW})-\sigma^{LO}(0,{0})}
\end{eqnarray}
Here $\mathcal{R}^{NLO}(\overline{c}_W,\overline{c}_{HW})$ is defined as the full NLO result, divided by the NLO SM piece plus the LO anomalous coupling pieces. The results for $W^+H$ and $ZH$ are presented in the figure. As might be expected from the inclusive $K$-factor the deviations are around $\pm 20\%$ depending on the position in the $(\overline{c}_W,\overline{c}_{HW})$ plane. Around $(-0.015,0.01)$ the NLO corrections suppress the result one would obtain if a LO prediction were used, and previous limits in this region of phase space (using the LO prediction) may be too optimistic. On the other hand, away from this region the corrections tend to be positive and will improve existing limits. We note that the region which corresponds to that in which our EFT calculation is valid intersects the region in which the impact of the NLO corrections is most rapidly changing. This suggests that using a flat $K$ factor to re-weight the anomalous coupling part of the calculation is not advisable. 

\begin{figure} 
\begin{center} 
\includegraphics[width=7cm]{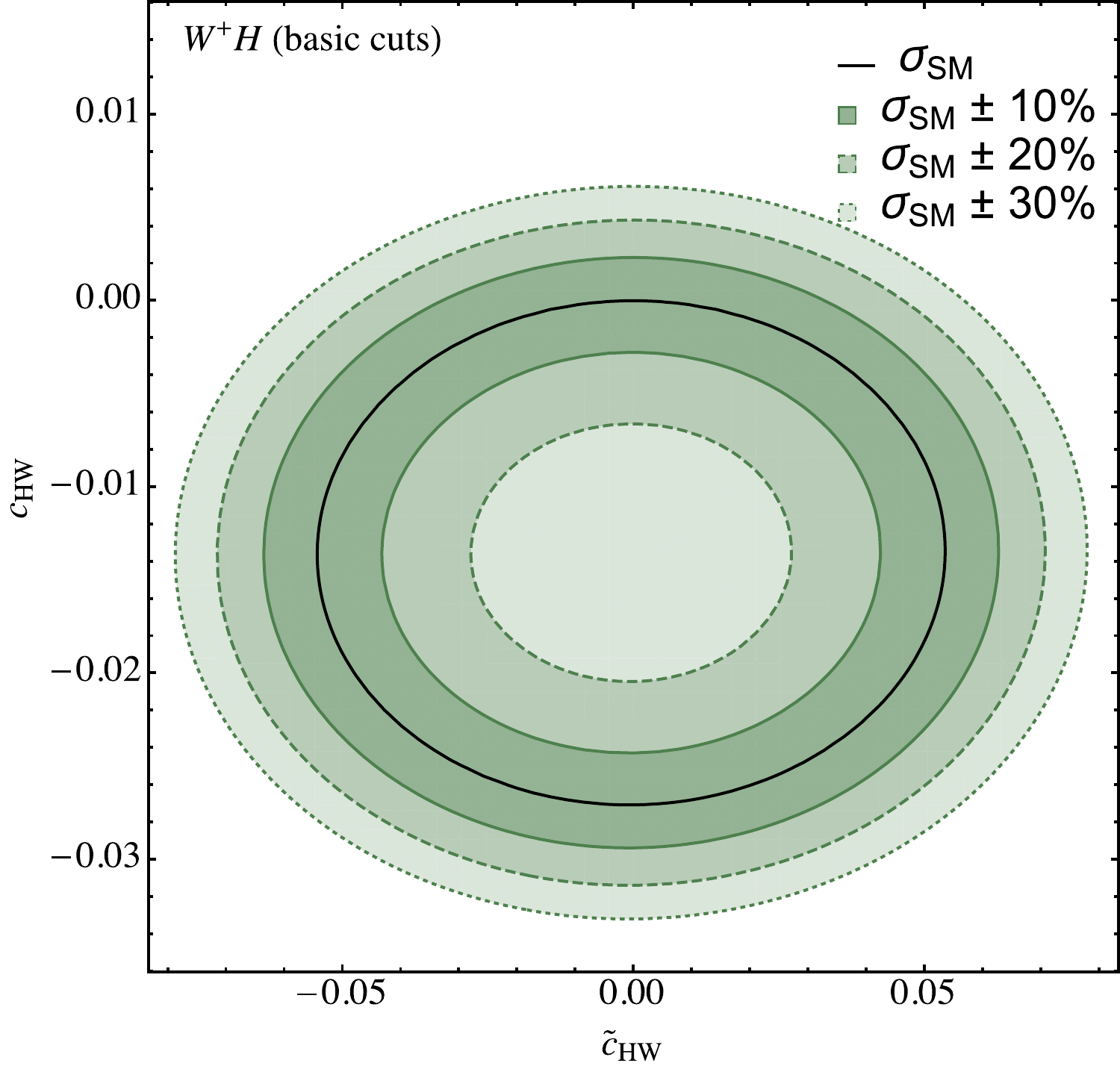} \includegraphics[width=7cm]{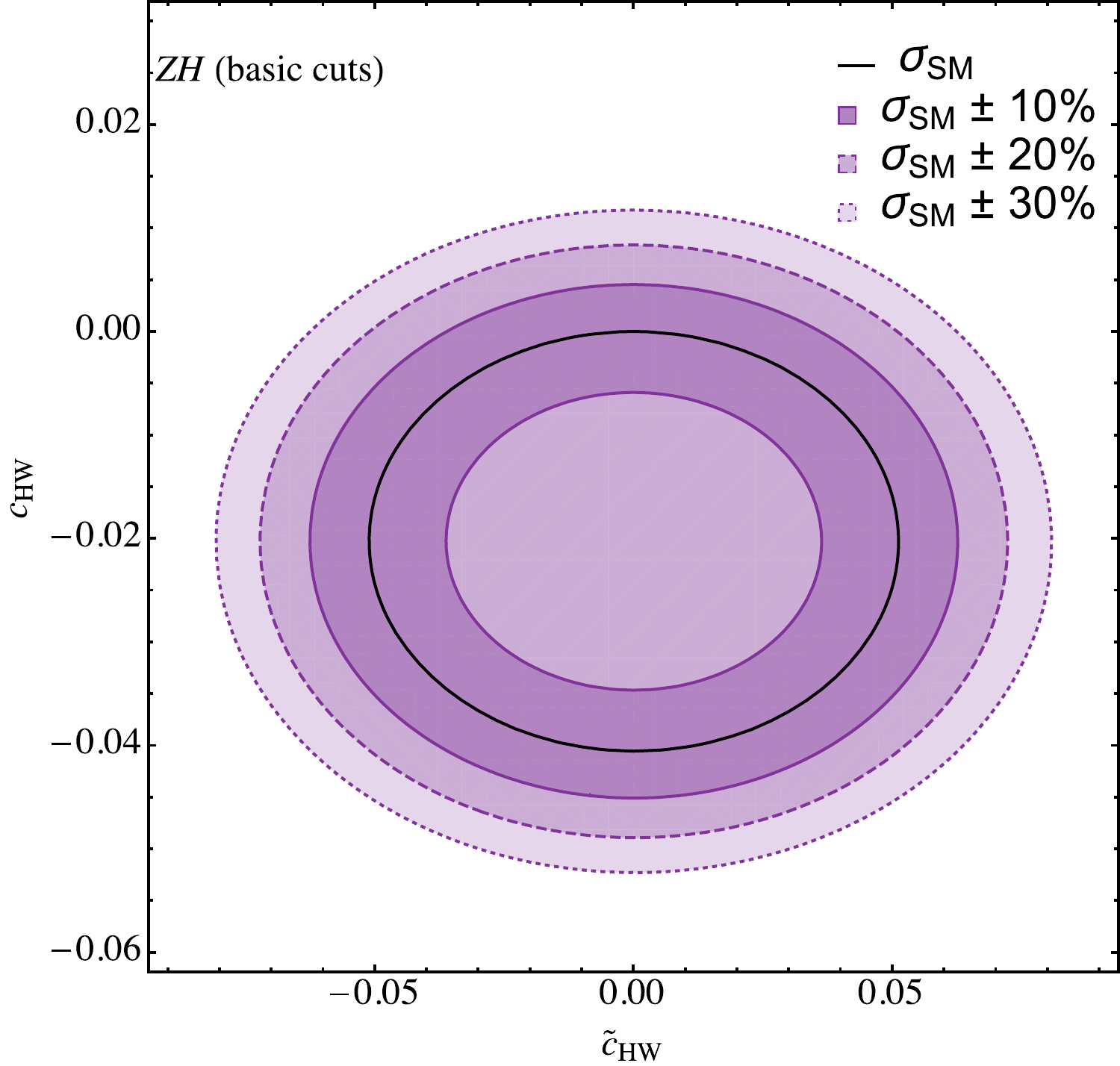} 
\caption{Contours representing deviations as a function of $\tilde{c}_{HW}$ and $\overline{c}_{HW}$ from the NLO SM prediction for $W^+H$ (left) and $ZH$ (right)  production at the 13 TeV LHC. The plots correspond to the basic cut selection. }
\label{fig:CPodd}
\end{center} 
\end{figure}

Finally in Figure~\ref{fig:CPodd} we present contours of constant cross section in the ($\tilde{c}_{HW},\overline{c}_{HW}$) plane, i.e. we study the impact of including CP odd operators. The CP odd operators do not interfere with the SM amplitude, so they first enter the cross section at $\mathcal{O}(\tilde{c}_{HW}^2)$. This can be seen in the figures via the $\tilde{c}_{HW} \leftrightarrow -\tilde{c}_{HW}$ symmetry in the figure. In order to have a relatively small deviation from the SM therefore requires a negative value of $\overline{c}_{HW}$ which can compensate for the positive definite correction arising from the CP odd operator. In Figure~\ref{fig:CPodd} we present results for the basic cuts only, however improved results could be obtained by optimizing the analysis to look at high $p_T^V$ observables. In addition angular distributions between final state particles are particularly sensitive to the CP structure of operators and represent a promising avenue of study. 

\subsection{NLO + Parton shower results}
We now turn to the presentation of the fixed order plus parton shower results making use of the MCFM/{\sc Powheg-Box} interface. The two processes considered are the production of the Higgs in association with a $Z$ or $W$ boson, where the Higgs decays to $b\bar{b}$ and the weak boson decays leptonically to $e$ or $\mu$.  Events were generated for some characteristic values of the EFT Wilson coefficients and showered/hadronised with {\sc Pythia8}. The decay of the Higgs was also performed by {\sc Pythia8}~\cite{Sjostrand:2007gs}, with the total rate normalised to the NLO production cross section times the branching fraction as calculated by e{\sc HDECAY}~\cite{eHDECAY} to NLO accuracy in both $\alpha_S$ and $\alpha_{EW}$. Event reconstruction and the implementation of basic selection cuts, summarised in Table~\ref{tab:cuts}, was performed using {\sc MadAnalysis5}~\cite{MA5} which makes use of {\sc Fastjet}~\cite{fastjet}. Benchmark EFT scenarios are selected to be within the allowed regions of recently performed global fits. In each case, an estimate of scale uncertainty is evaluated by varying the renormalisation and factorisation scales between half and twice the central scale, which is the invariant mass of the $VH$ system. This is combined in quadrature with the usual Monte Carlo statistics uncertainty arising from the finite number of events generated. NLO and LO samples were generated with the CTEQ10 and CTEQ6L1 PDF sets respectively and PDF uncertainties were not estimated. 

\begin{table}
\begin{center}
    \begin{tabular}{|c|c|}
        \hline
        \multicolumn{2}{|c|}{Process}\tabularnewline \hline
        $H \,Z\,\to b\bar{b}\,\ell^+\ell^-$&
        $H \,W\to b\bar{b}\,\ell\nu$\tabularnewline\hline
        \multicolumn{2}{|c|}{Jets}\tabularnewline \hline
        \multicolumn{2}{|c|}{$k_T$ algorithm: $\Delta R$=0.4, $p_T > 25$ GeV \& $\eta_b < 2.5$} \tabularnewline \hline
        \multicolumn{2}{|c|}{Cuts}\tabularnewline \hline
        \multicolumn{2}{|c|}{2 $b$-jets, $p_T > 25$ GeV, $\eta_b < 2.5$} \tabularnewline \hline
        1 lepton, $\ell^\pm$ ($e$ or $\mu$)&
        2 leptons, $\ell^+,\,\ell^- $ ($e$ or $\mu$)\tabularnewline \hline
        \multicolumn{2}{|c|}{$p_T^{\ell} < 25$ GeV, $|\eta_{\ell}| < 2.5$} \tabularnewline \hline
    \end{tabular}
    \end{center}
    \caption{\label{tab:cuts} Table summarizing the selection cuts performed on events in the two vector boson associated production modes.}
\end{table}

\subsubsection{Gluon initiated contribution to $HZ$\label{s:ggHZ}}
\begin{figure} 
\begin{center} 
\includegraphics[width=4.9cm]{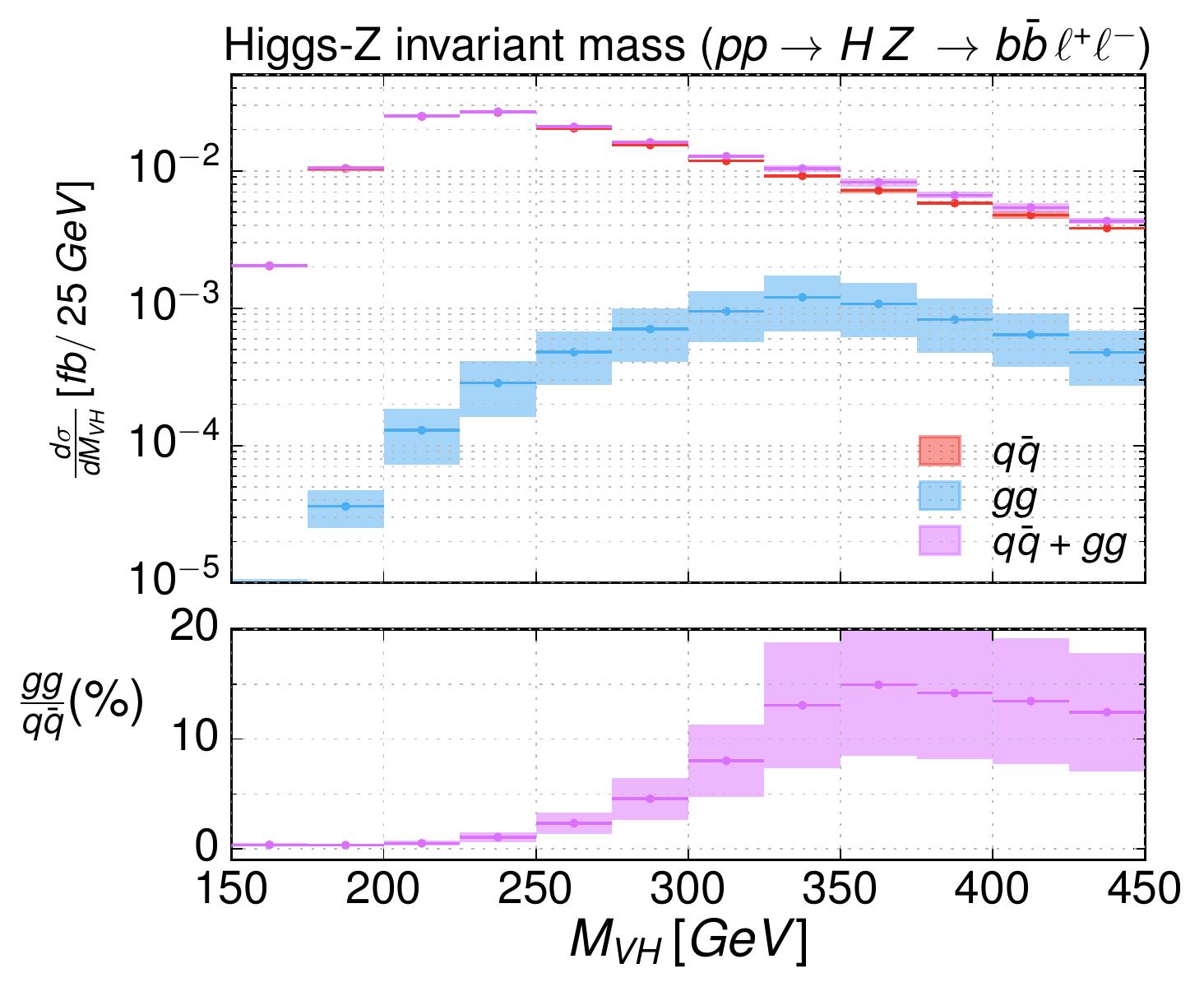}
\includegraphics[width=4.9cm]{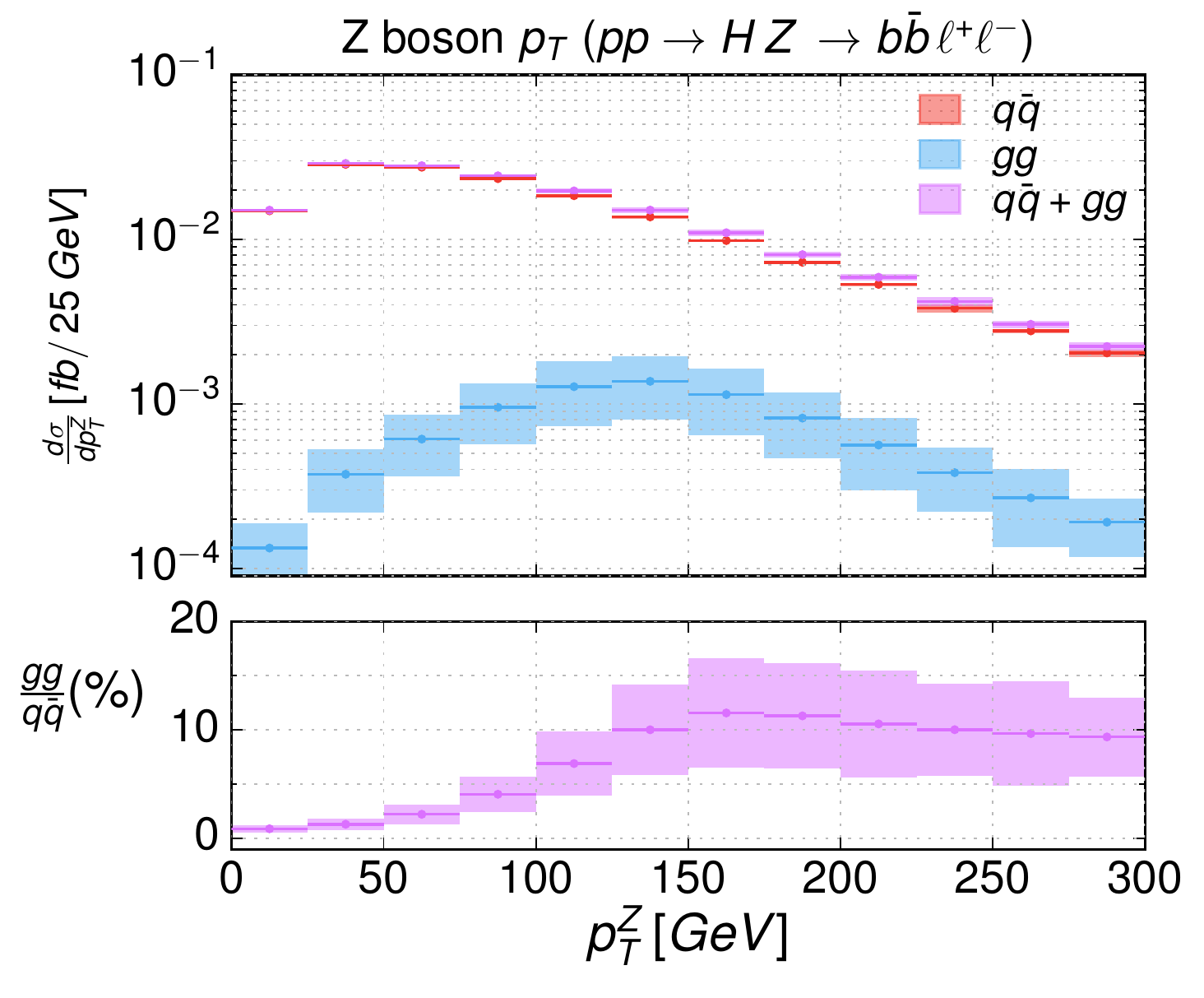}
\includegraphics[width=4.9cm]{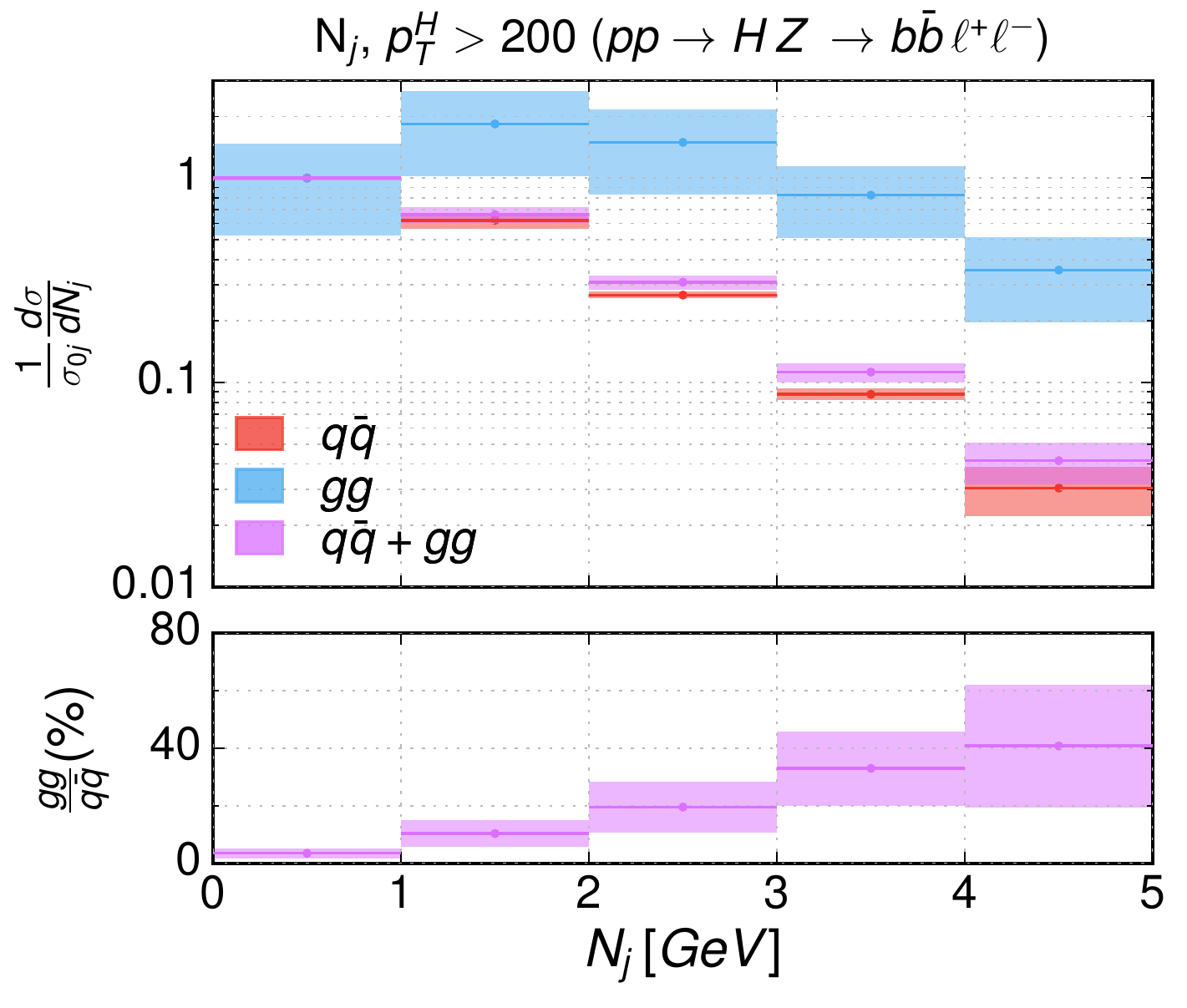}
\caption{Impact of the $gg$-initiated contribution to $pp\to H Z \to b\bar{b}\,\ell^+\ell^-$. Upper panels show differential distributions in, from left to right, the invariant mass of the $H Z$ system, the $p_T$ of the Z boson and the $N_{\text{jets}}$ distribution normalised to the 0-jet bin after a cut on $p_T^{Z}$ of 200 GeV. Lower panels show the ratio of the $gg$- and $q\bar{q}$-initiated contributions.
\label{fig:gg_NLOPS}}
\end{center} 
\end{figure}

In order to highlight the importance of the $gg$-initiated contribution to $ZH$ production, a sample was generated separately and compared to the pure $q\bar{q}$-initiated piece at NLO. In general, due to the $2m_t$ thresholds, the kinematics of the box configuration will prefer a significantly harder region of phase space than the Drell-Yan like topology and, if it is not adequately taken into account, could show up as a fictitious EFT-like signal. Figure~\ref{fig:gg_NLOPS} overlays the two contributions in several differential distributions, showing the relative size of the would-be `signal' that one may observe if the $gg$ piece were not factored into the SM prediction. The contribution of the sub-process to the inclusive cross section is minor, of order 3--4\%, but as it populates a high $p_T$ region, where the SM cross section is also quite small, the $gg$ contribution can show up as an $\mathcal{O}(10-15\%)$ effect in the tails of differential distributions, mimicking a potential EFT-like signature. The effect of this contribution on the $N_j$ distribution is even more striking, given the increased emission probabilities of the initial state, reaching around 40\%. In Sections~\ref{s:EFTVH}, this contribution is taken into account as part of the SM prediction for $ZH$ associated production.
The final panel in Figure~\ref{fig:gg_NLOPS} (and similar panels in future figures) encapsulates the emission of radiation in the form of the number of jets in the process. These jets can arise from either the matrix element or the subsequent parton shower. Since the matrix elements can provide at most one additional jet at NLO, 
the parton shower provides the additional jet multiplies beyond those of the NLO matching, which for these processes corresponds to $N_{j} > 3$ ($q\overline{q}$), and $N_j > 2$ $(gg)$ (and hence these rates are subject to larger theoretical uncertainties.) 

\subsubsection*{LO +PS vs NLO +PS \label{s:kfact}}
\begin{figure} 
\begin{center} 
    \begin{tabular}{l}
        \hline
        $pp\to HZ\to b\bar{b}\ell^+\ell^-$\\[0.2ex]
        \hline\\
\includegraphics[width=4.9cm]{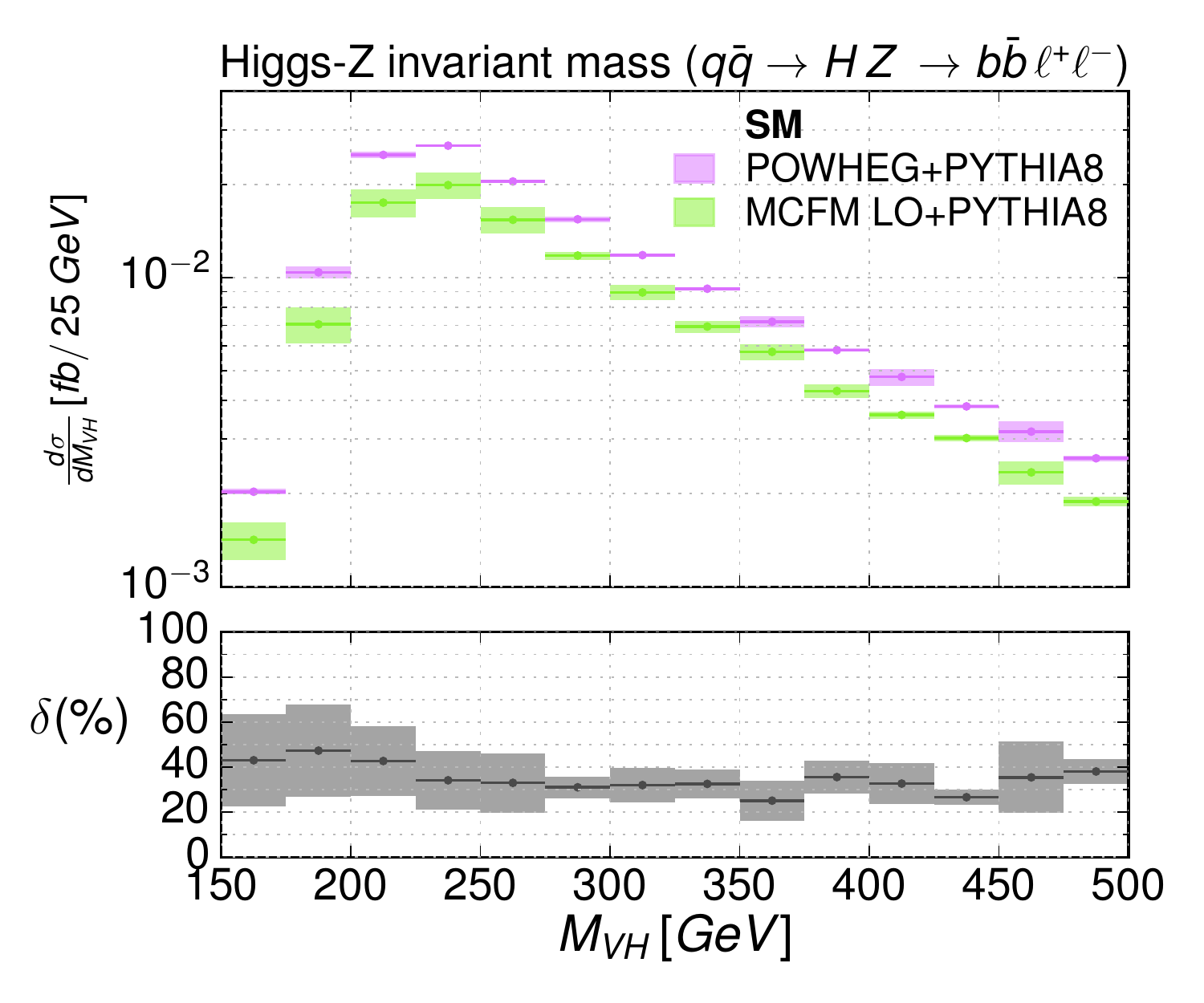} 
\includegraphics[width=4.9cm]{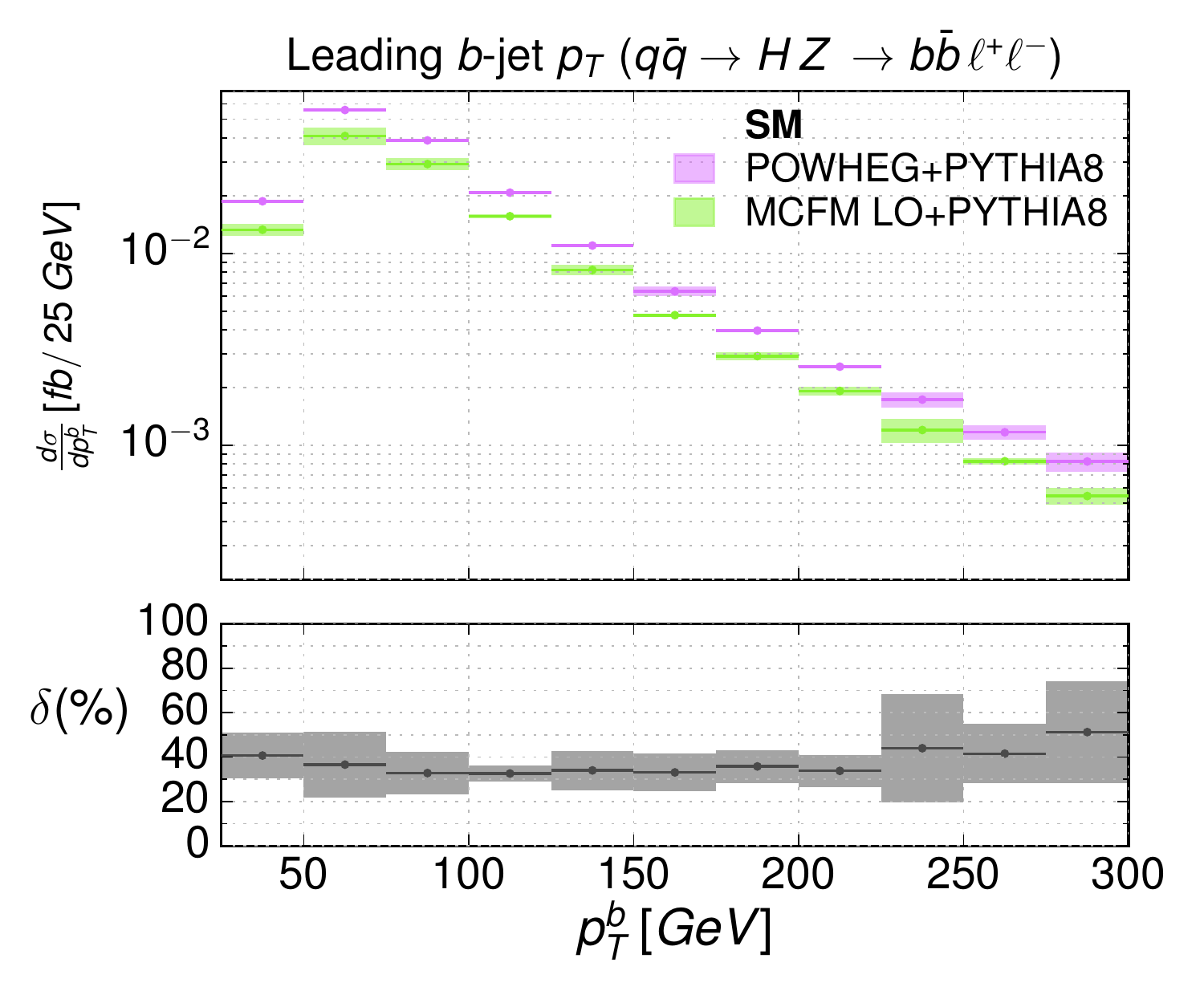}
\includegraphics[width=4.9cm]{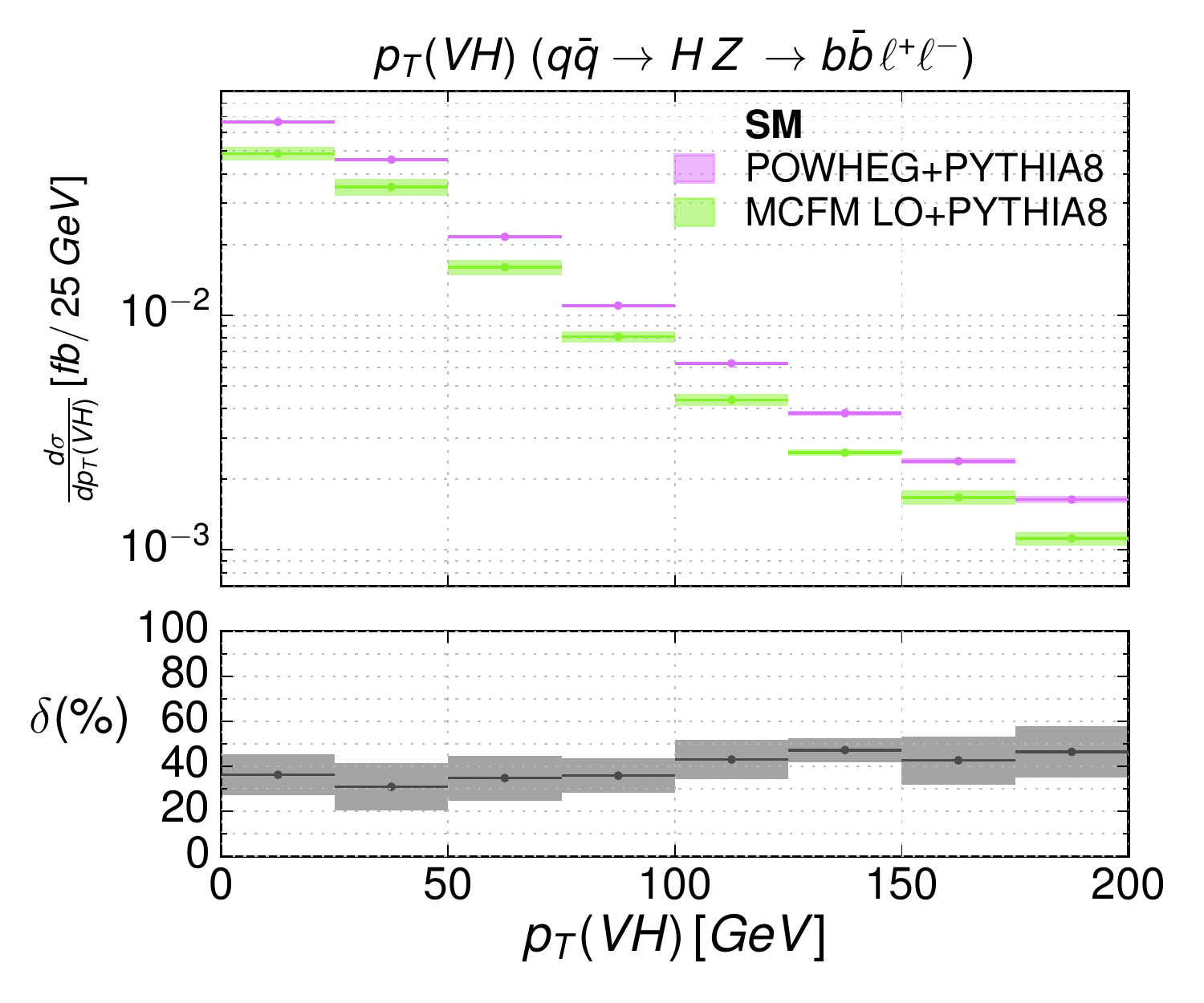}\\
\includegraphics[width=4.9cm]{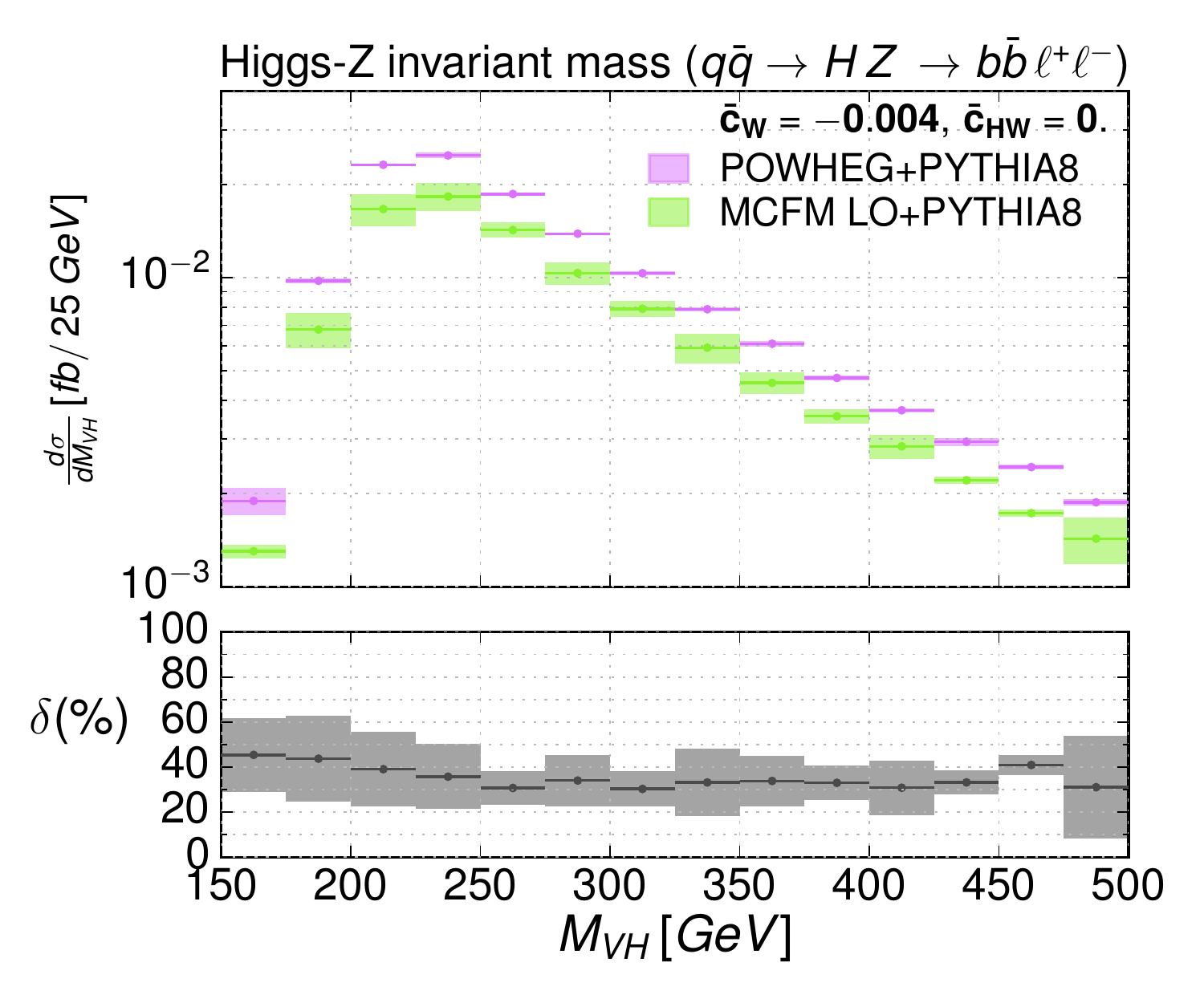} 
\includegraphics[width=4.9cm]{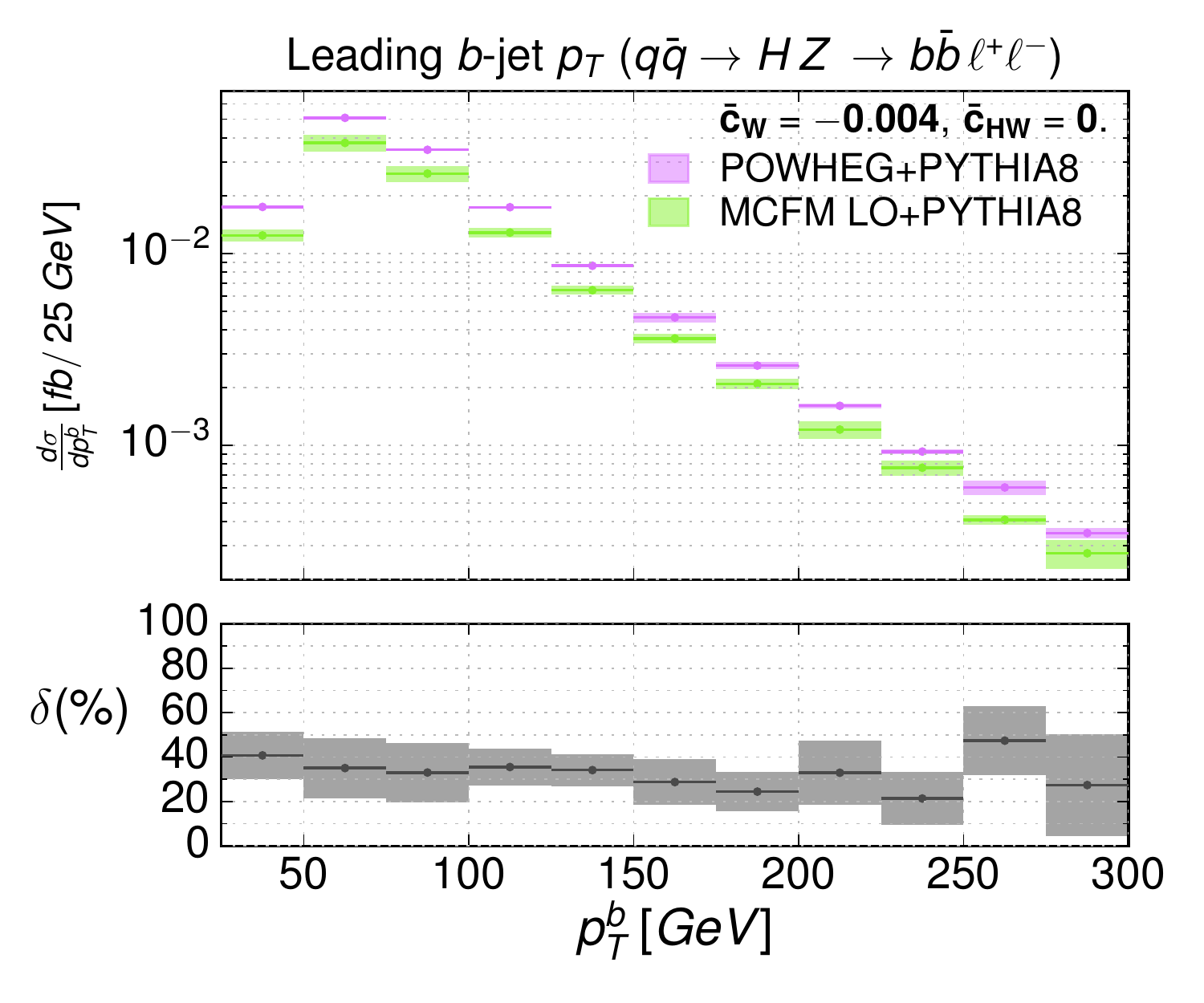}
\includegraphics[width=4.9cm]{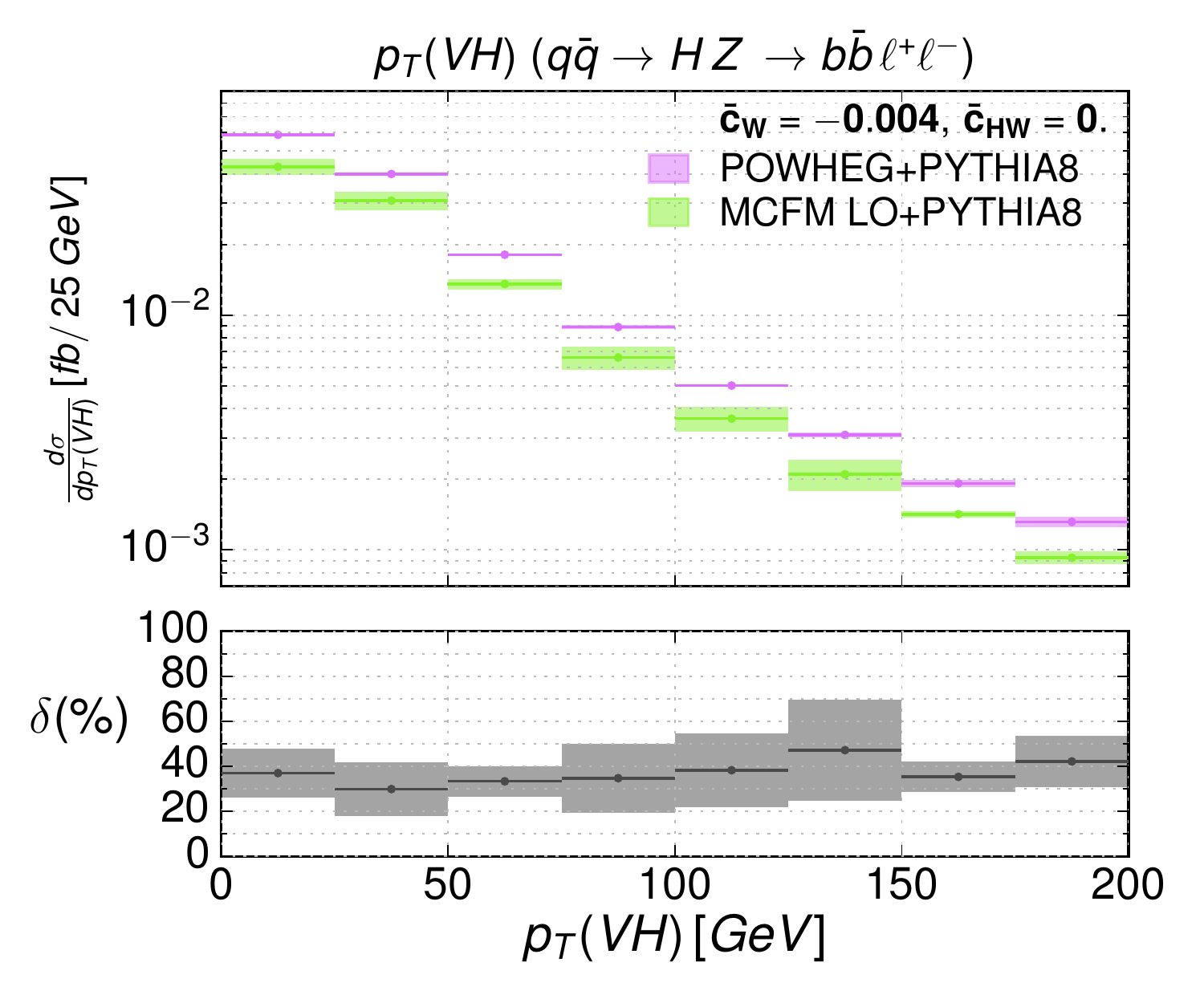}\\
        \hline
        $pp\to HW^+\to b\bar{b}\ell^+\nu$\\[0.2ex]
        \hline\\
\includegraphics[width=4.9cm]{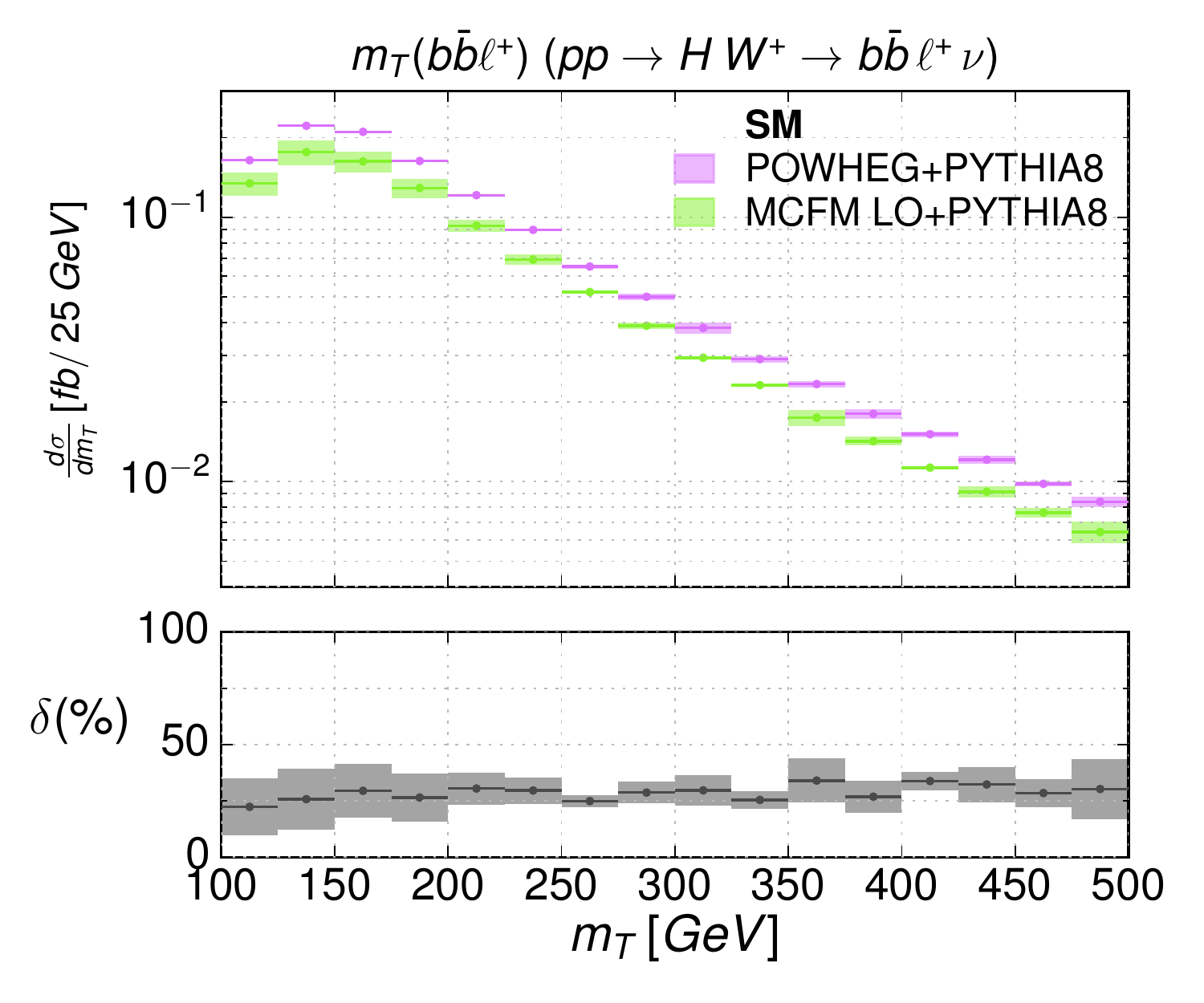} 
\includegraphics[width=4.9cm]{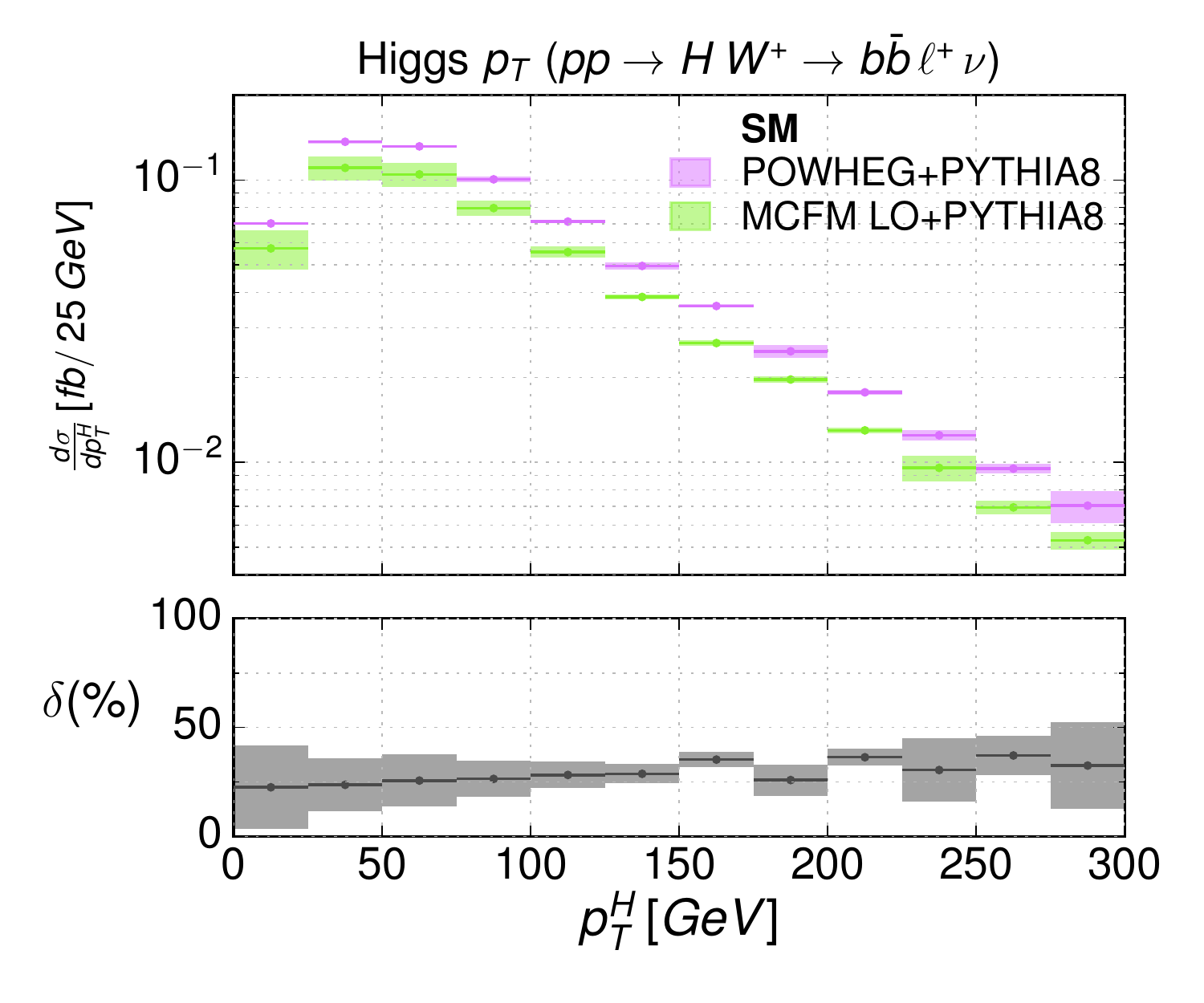}
\includegraphics[width=4.9cm]{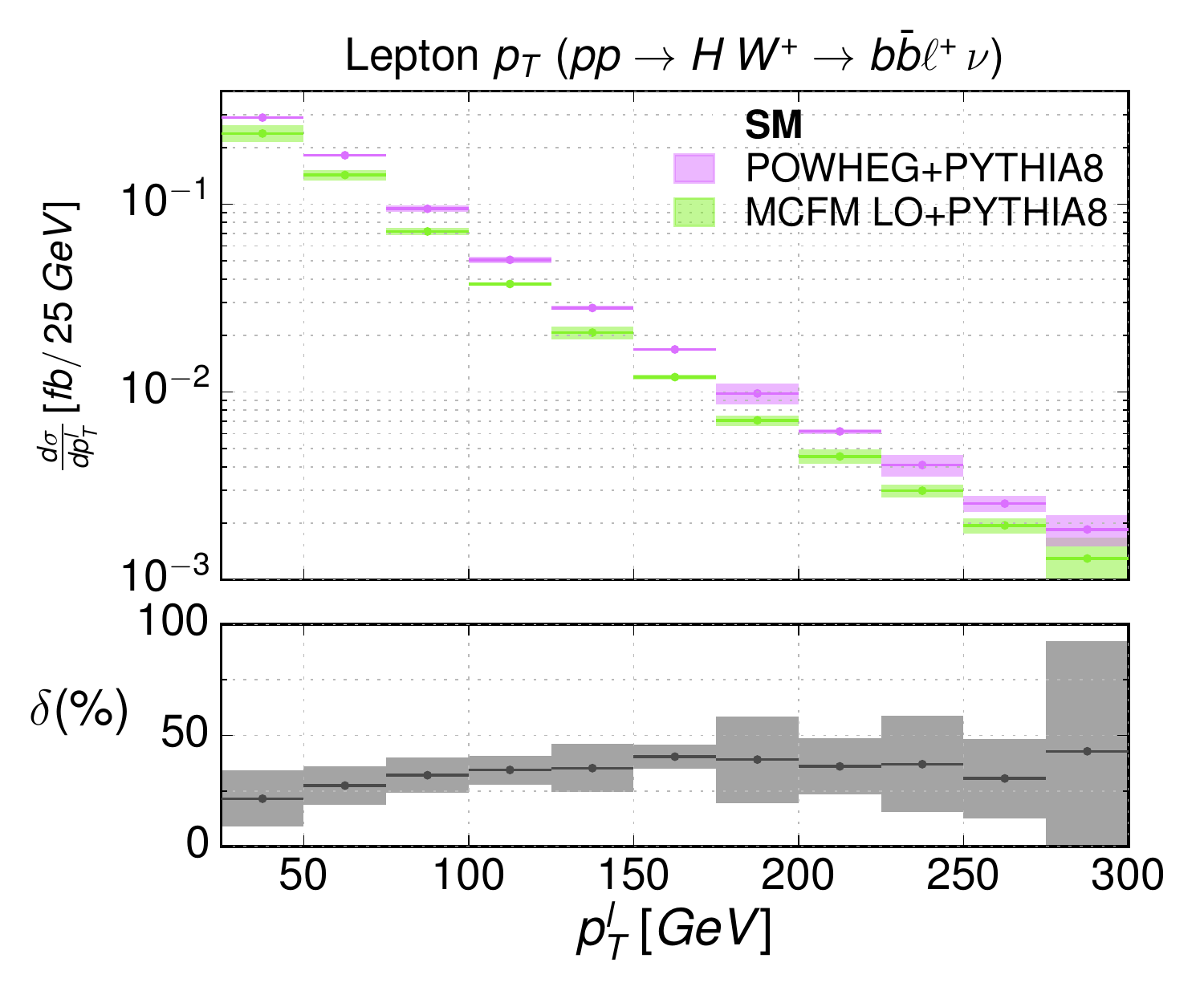}\\
\includegraphics[width=4.9cm]{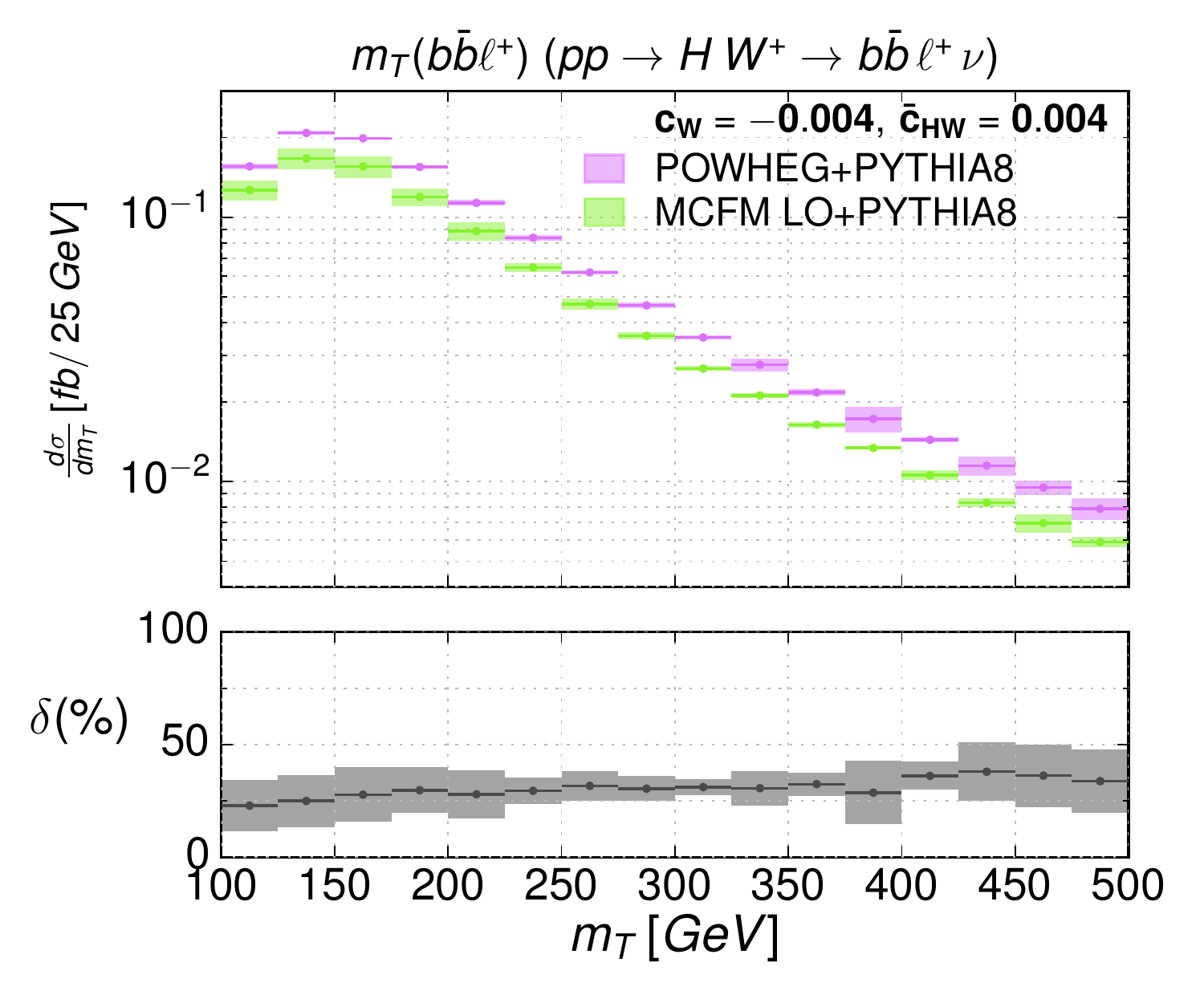} 
\includegraphics[width=4.9cm]{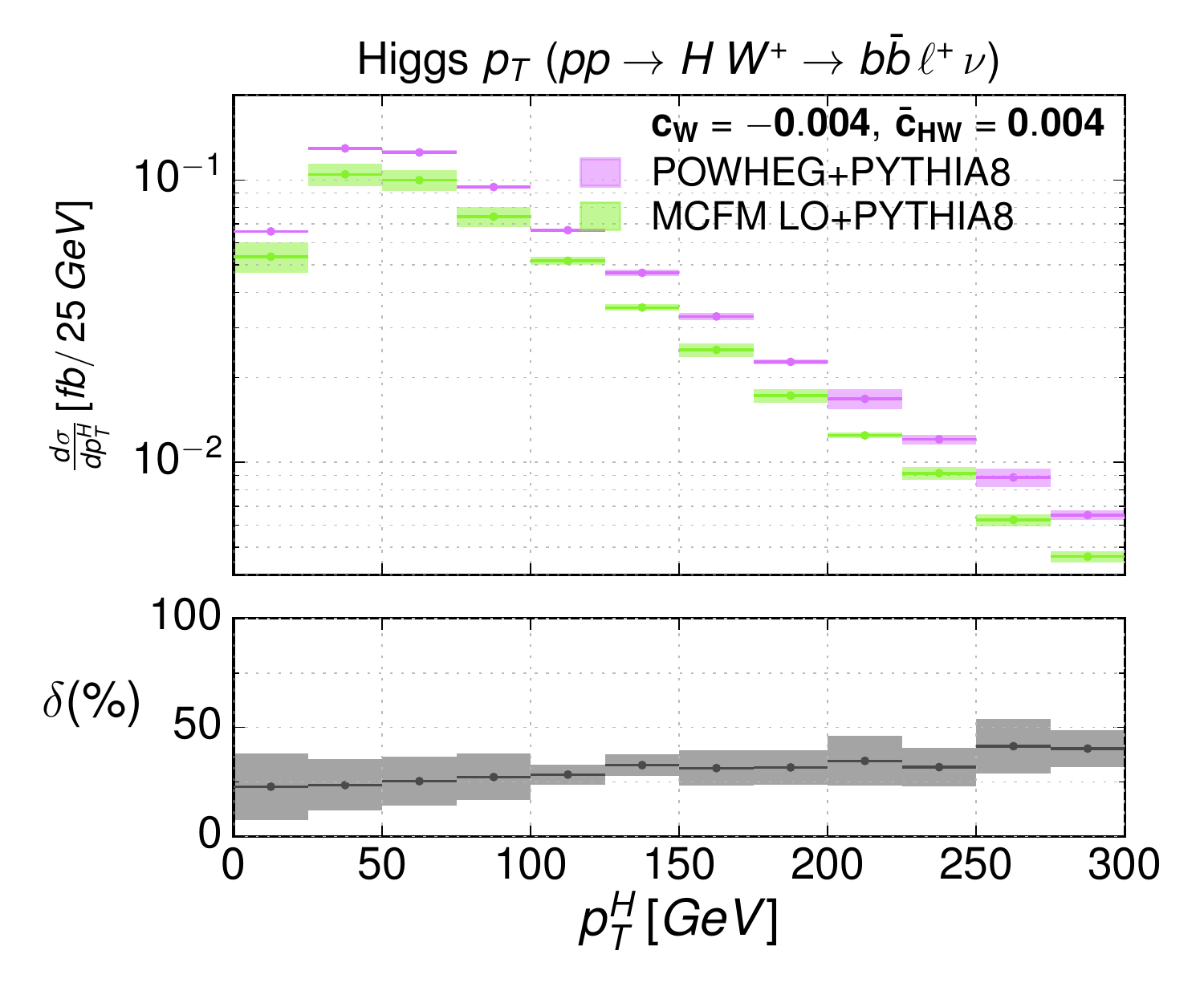}
\includegraphics[width=4.9cm]{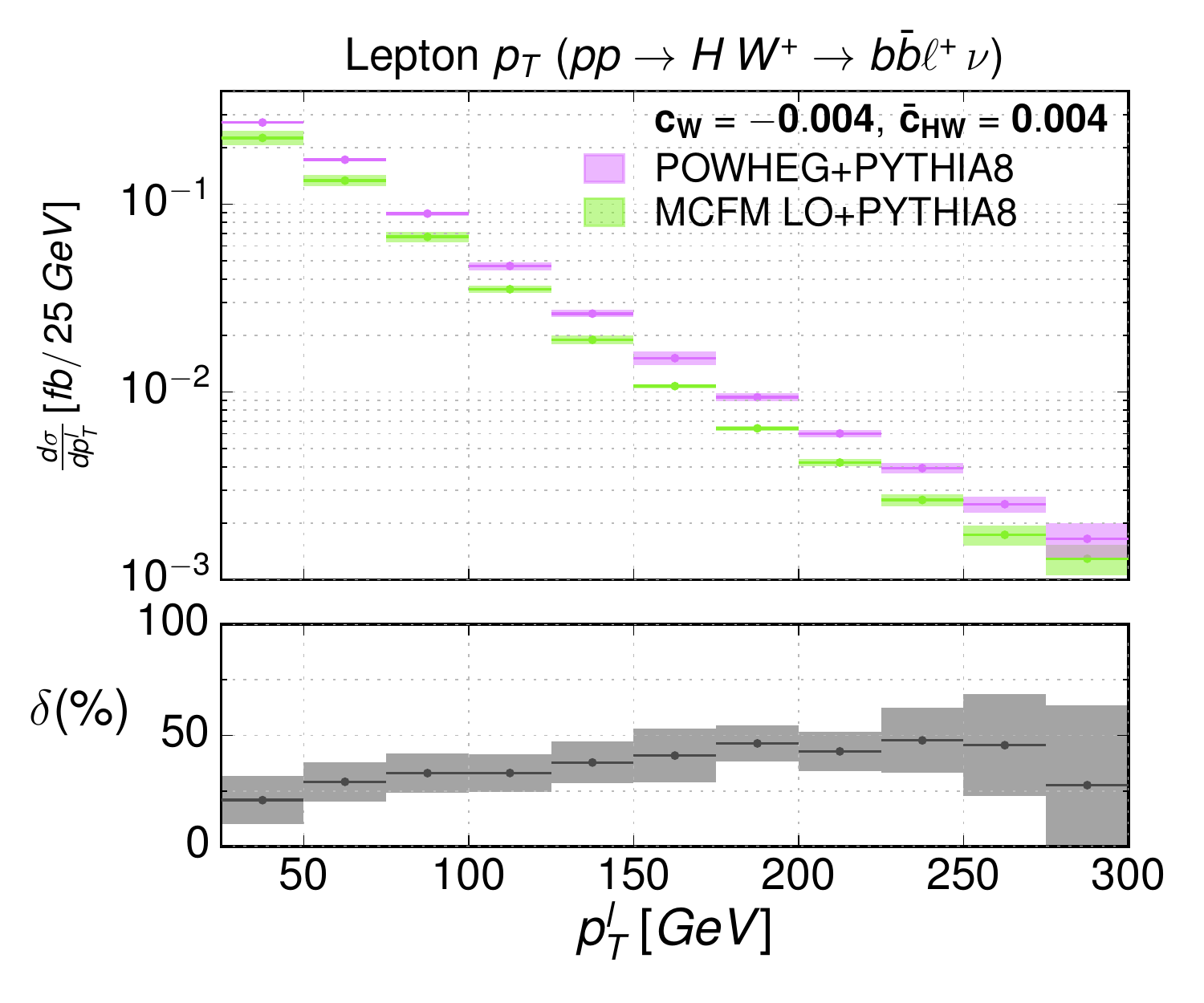}
    \end{tabular}
\caption{Comparison of differential distributions in $pp\to HZ\to b\bar{b}\ell^+\ell^-$ and $pp\to HW^+\to b\bar{b}\ell^+\nu$ between the LO MCFM and NLO {\sc Powheg} implementations, both showered/hadronized with {\sc Pythia8}. For the $ZH$ case, the comparison is made in the SM (\emph{upper row}) and the benchmark point of $\bar c_W=0.004$ (\emph{lower row}), while in the $WH$ case, the comparison is made in the SM (\emph{upper row}) and the benchmark point of $\bar c_W=-\bar c_{HW}=-0.004$ (\emph{lower row}). \label{fig:kfact}}
\end{center} 
\end{figure}

To asses the impact of taking higher order effects into account, we now compare the new implementation at NLO to a LO one in MCFM, post showering and hadronisation. Figure~\ref{fig:kfact} depicts a selection of differential distributions in the SM and for one of our benchmark points of $\overline{c}_W=0.004$ (see discussion in Section~\ref{s:EFTVH}) for $HZ$ and $HW$ production respectively. Here the predominant effect is that of a relatively flat $K$-factor that is not sensitive to the presence of the new EFT interactions which, in any case, are colour neutral. The rightmost distributions in the upper half of Figure~\ref{fig:kfact} show the $p_T$ of the $ZH$ system and are therefore sensitive to the `kick' that it receives from additional radiation. We see a mild rise in the tail between the NLO case that we would expect given that it captures the full phase space of one additional emission compared to the LO case which remains within the soft and collinear approximation of the parton shower. 

\subsubsection{EFT effects\label{s:EFTVH}}
\begin{figure} 
\begin{center} 
\includegraphics[width=4.9cm]{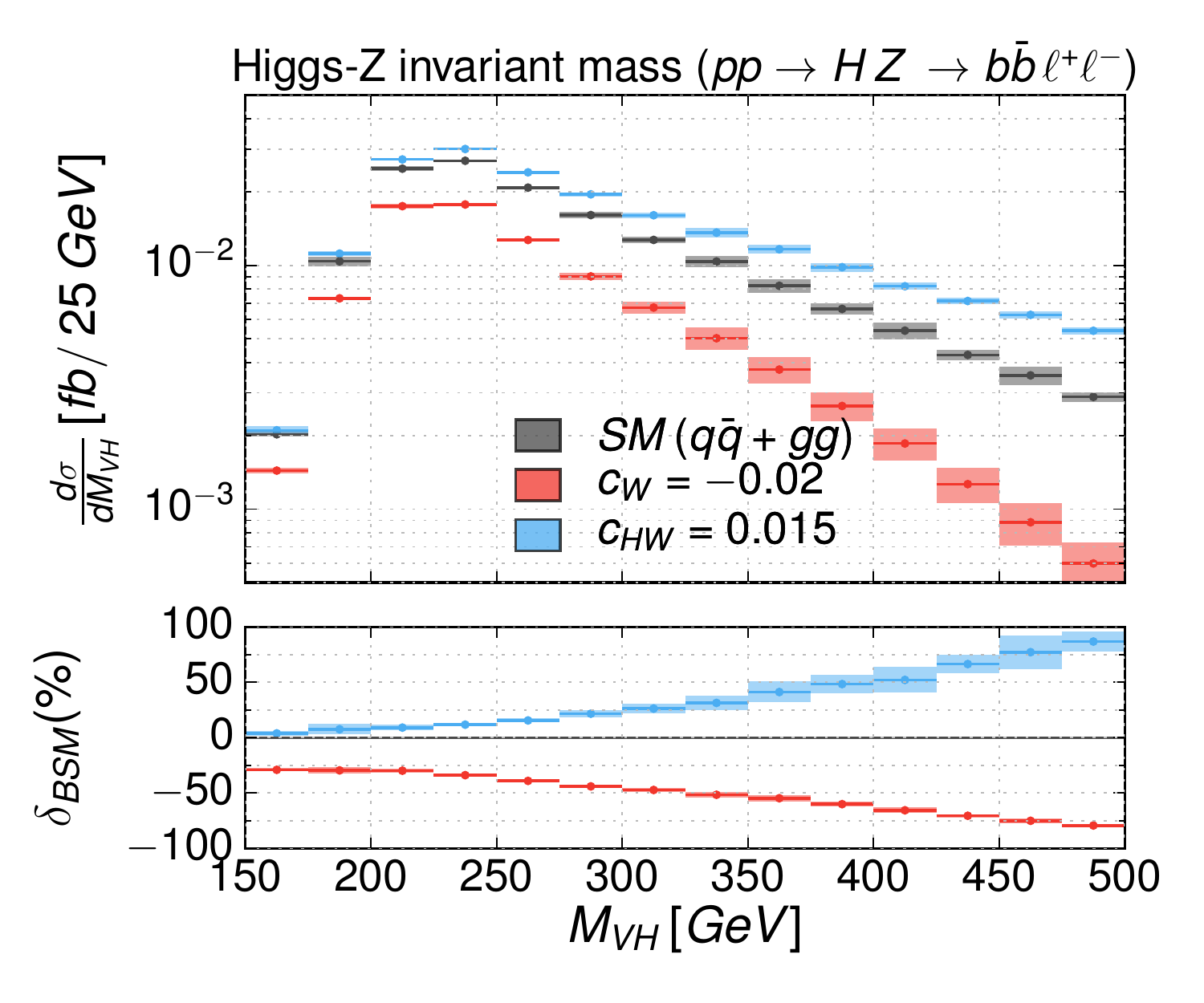} 
\includegraphics[width=4.9cm]{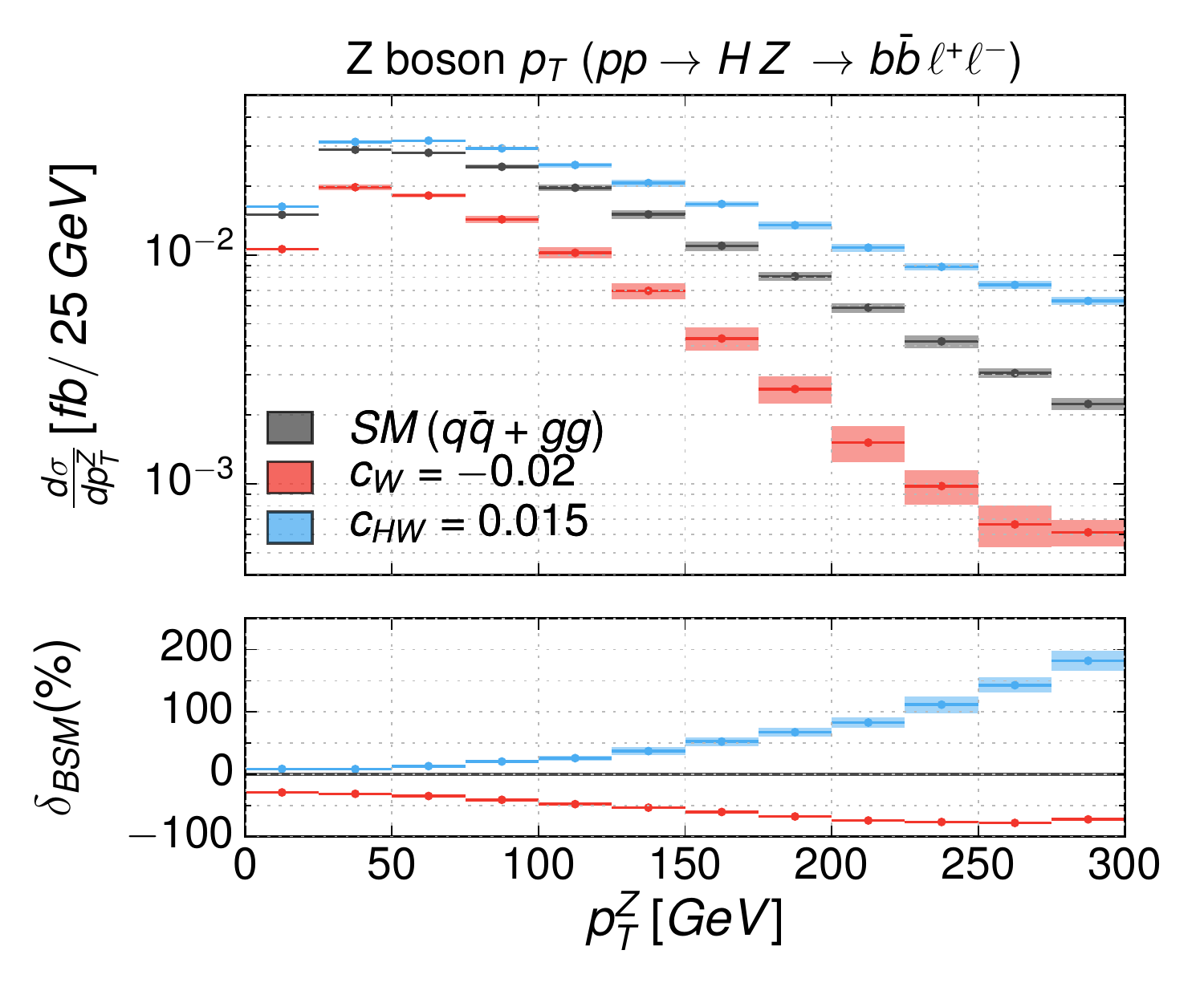}
\includegraphics[width=4.9cm]{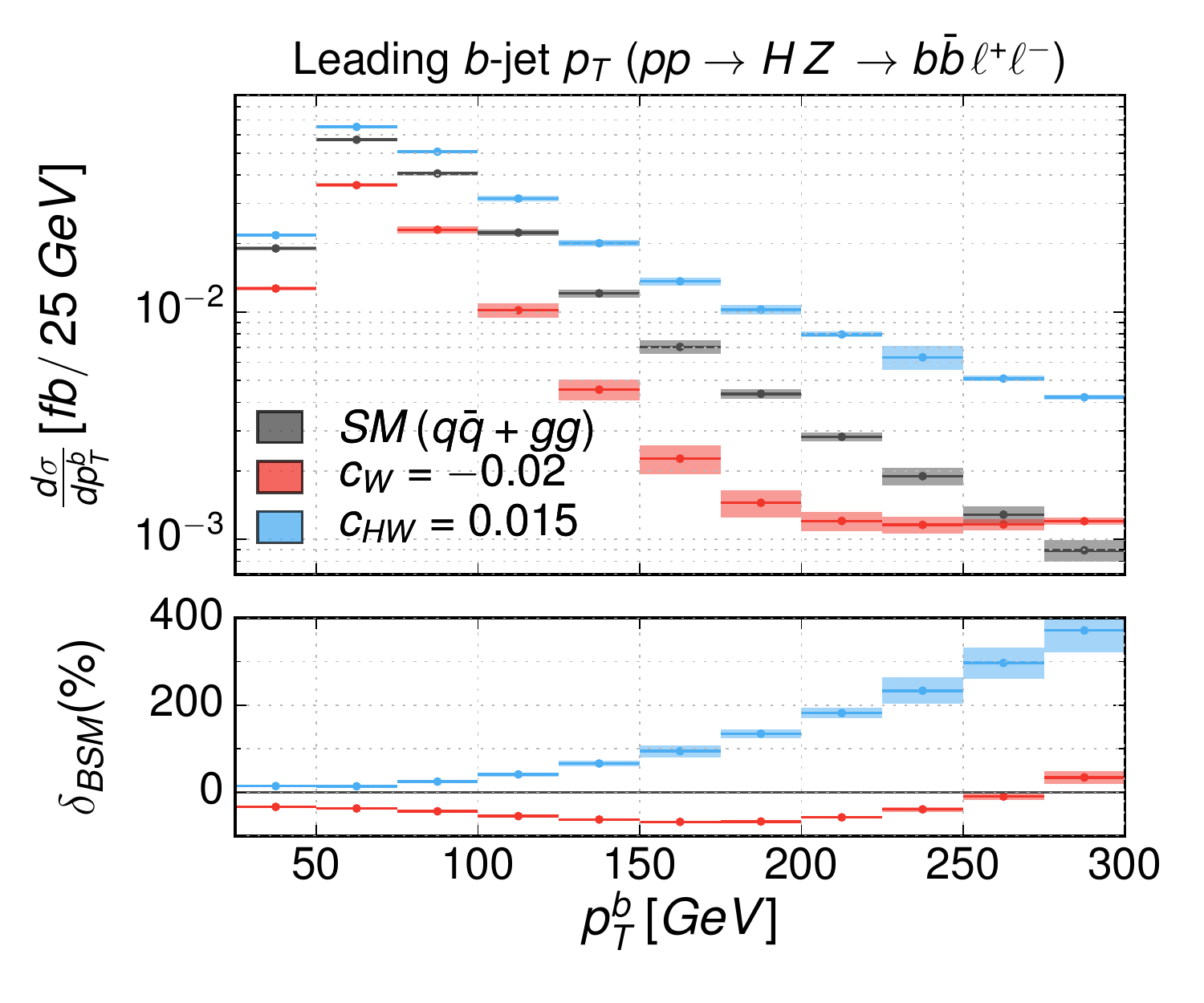} 
\caption{\label{fig:ZH_NLOPS_lg}  Comparison of the SM prediction with values of the Wilson coefficients of $\bar c_W=-0.02$ and $\bar c_{HW}=0.015$. From left to right, the differential cross sections with respect to the Higgs-Z invariant mass, $m_{VH}$; Z-boson transverse momentum, $p_T^Z$; and the leading $b$-jet transverse momentum, $p_T^{b1}$, are shown in the upper panels, with the percentage deviation of the EFT benchmarks from the SM prediction, $\delta_{BSM}$ shown in the lower panels.}
\end{center} 
\end{figure}

We now turn to examining the effect of switching on one or more of the previously defined Wilson coefficients that affect associated production in both the $ZH$ and $WH$ channels. We limit ourselves to the $\overline{c}_W$ and $\overline{c}_{HW}$ coefficients as they are sufficient to capture the additional non SM-like momentum dependence brought about by dimension-6 operators. Figure~\ref{fig:gg_NLOPS} displays a number of characteristic differential distributions evaluated using the values of $\overline{c}_W =-0.02$ and $\overline{c}_{HW}=0.015$, which saturate the bounds set by the most recent global fits~\cite{withJohn}. In general, very large effects are expected for such sizable values of the coefficients and considering the discussion in Section~\ref{sec:FO}, the validity of such an EFT description is called into question in the phase space regions where the BSM effects are important. Considering Figure~\ref{fig:ZHcwchw}, it is clear that the values of the coefficients lie well outside of the regions in which the quadratic piece of the EFT contribution makes up less than 10\% of the overall contribution. The $p_T^{b}$ distribution in Figure~\ref{fig:ZH_NLOPS_lg}, for example, highlights very clearly the onset of a breakdown of the EFT in the high $p_T$ tail, where the relative contribution of the $\overline{c}_W=-0.02$ point changes sign, suggesting the dominance of the $(\overline{c}_W/\Lambda)^2$ term. Therefore, although these values of Wilson coefficients are technically `allowed', the evidence in this section as well as in Section~\ref{sec:FO}, suggests that we are not yet at a point in which the sensitivity of experiments can provide meaningful information about the coefficients affecting this Higgs production process. 

Since we have not included effects from dimension 8 operators, which may be as large as the aforementioned squared EFT contributions, we prefer to present results using more conservative values of the coefficients, where the EFT interpretation is better motivated. These values are chosen from the criteria delineated in Section~\ref{sec:FO}, \emph{i.e.}, the requirement that the squared terms do not make up more than 10\% of the overall contribution. This leads us to choose $|\bar c_{W}|,|\bar c_{HW}|=0.004$. Our two benchmark points derived from this are $\bar c_W=0.004$ and $\bar c_W=-\bar c_{HW}=-0.004$. The relationship imposed in the latter choice is motivated by the results of previous works that calculated the low energy EFT coefficients predicted by a number of UV scenarios~\cite{withJosemi}. In the Two-Higgs Doublet Model, for example, this relationship is always satisfied at the matching scale. From a phenomenological perspective, this relation is also special because it corresponds to the elimination of one of the two momentum structures, $g_{hvv}^{(2)}$, present in the extended Higgs-gauge vertices (see Table~\ref{trans} and discussion in Section~\ref{sec:EFT}). 

\begin{figure} 
\begin{center} 
\includegraphics[width=4.9cm]{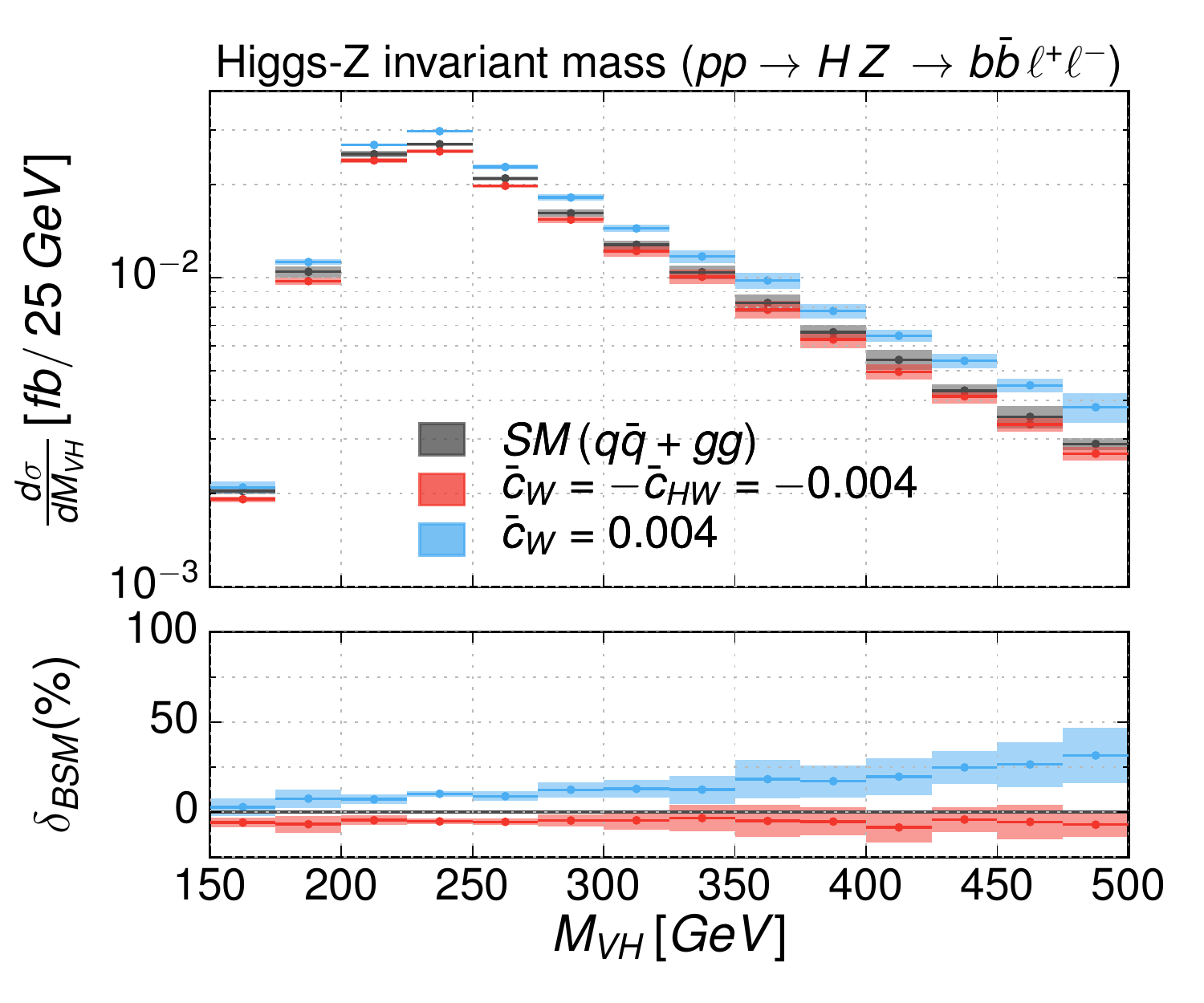} 
\includegraphics[width=4.9cm]{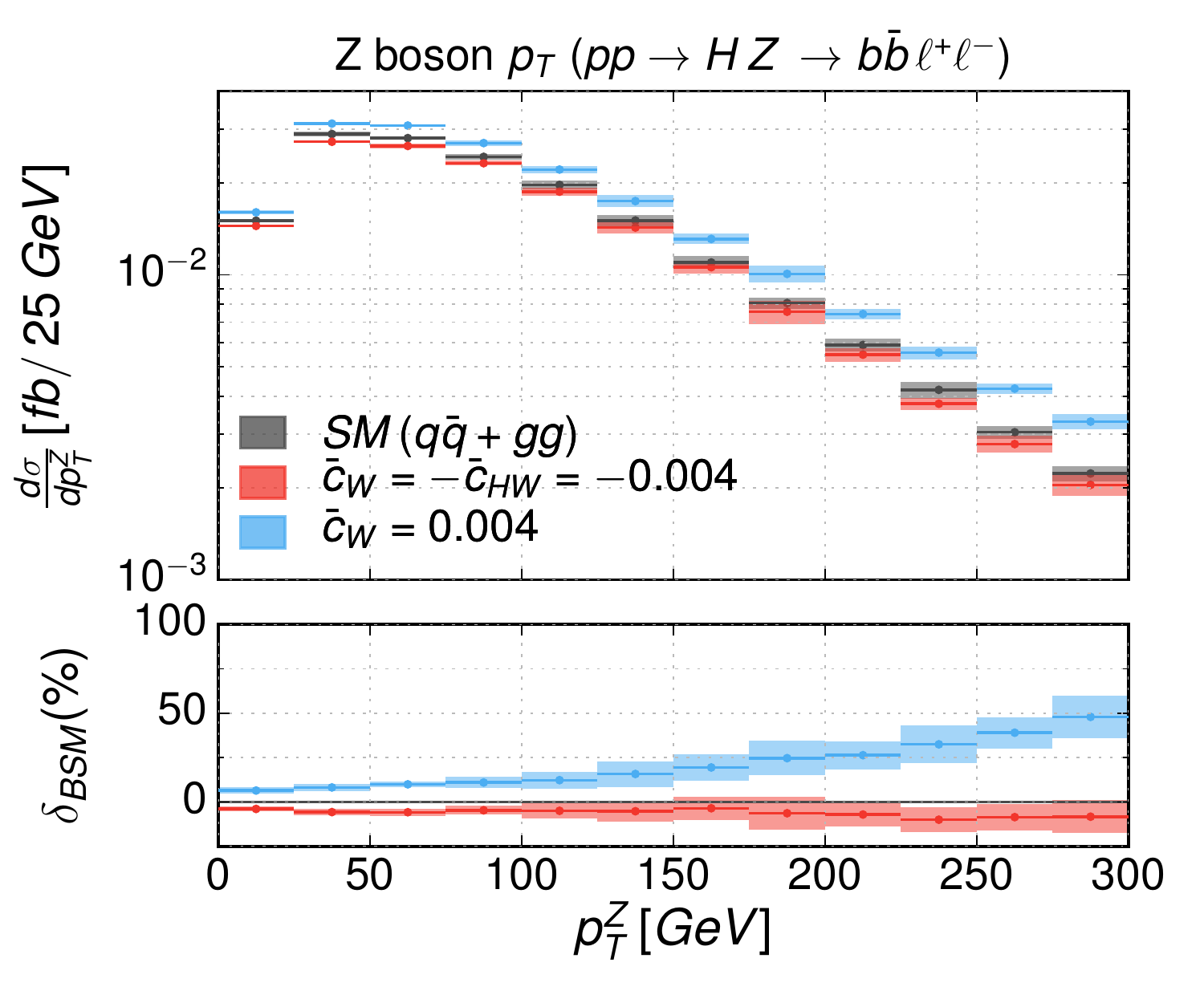} 
\includegraphics[width=4.9cm]{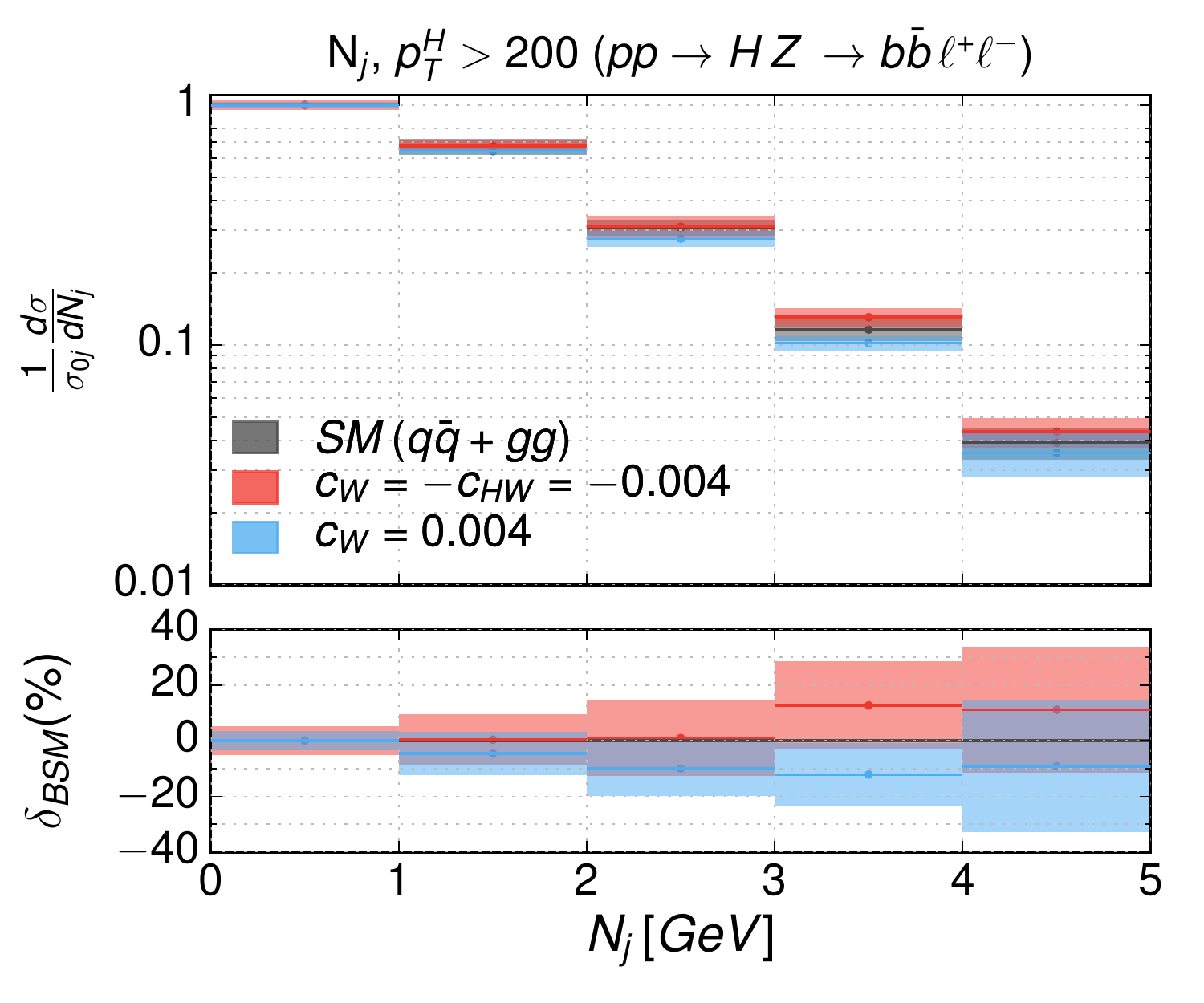}\\
\includegraphics[width=4.9cm]{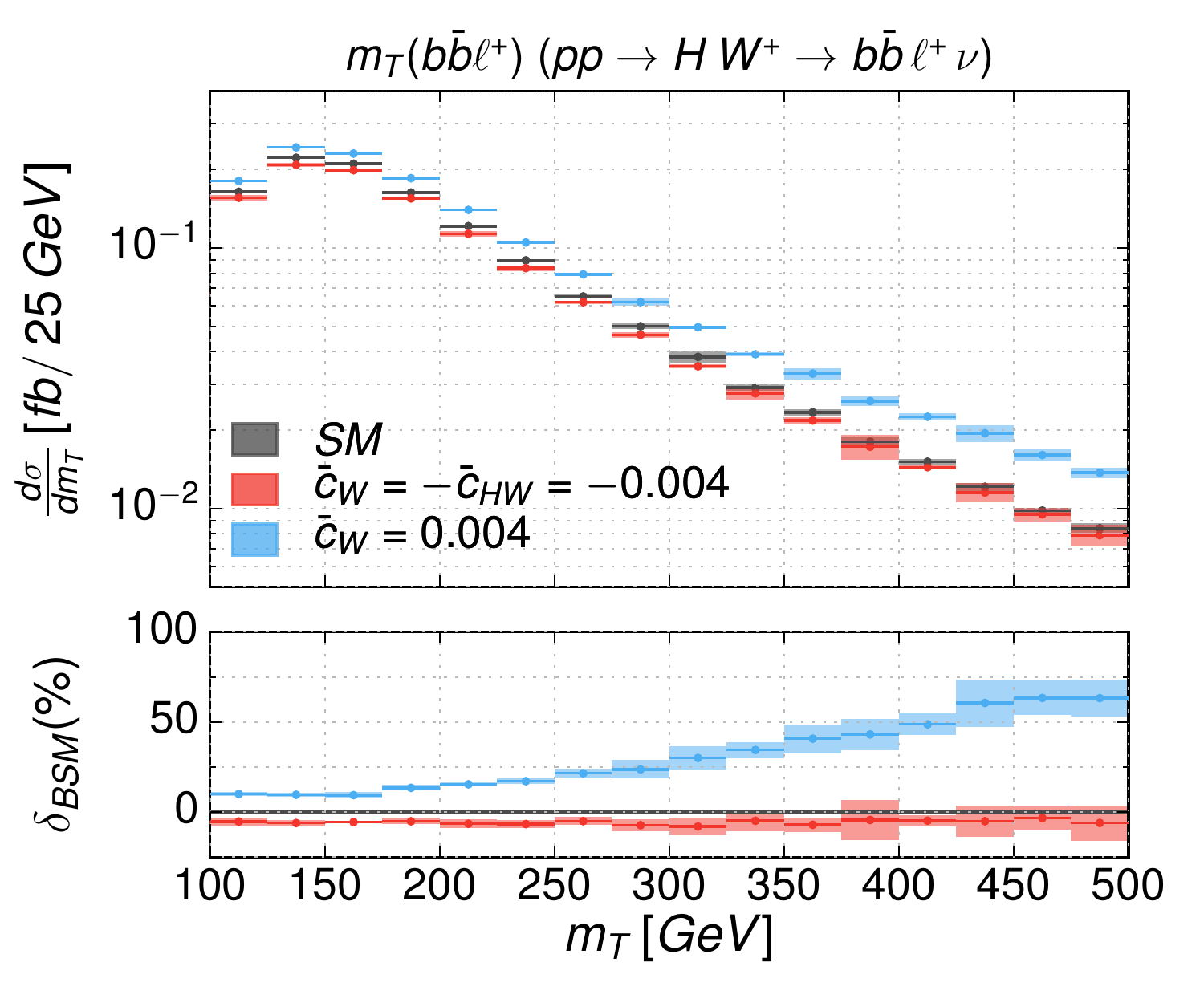}
\includegraphics[width=4.9cm]{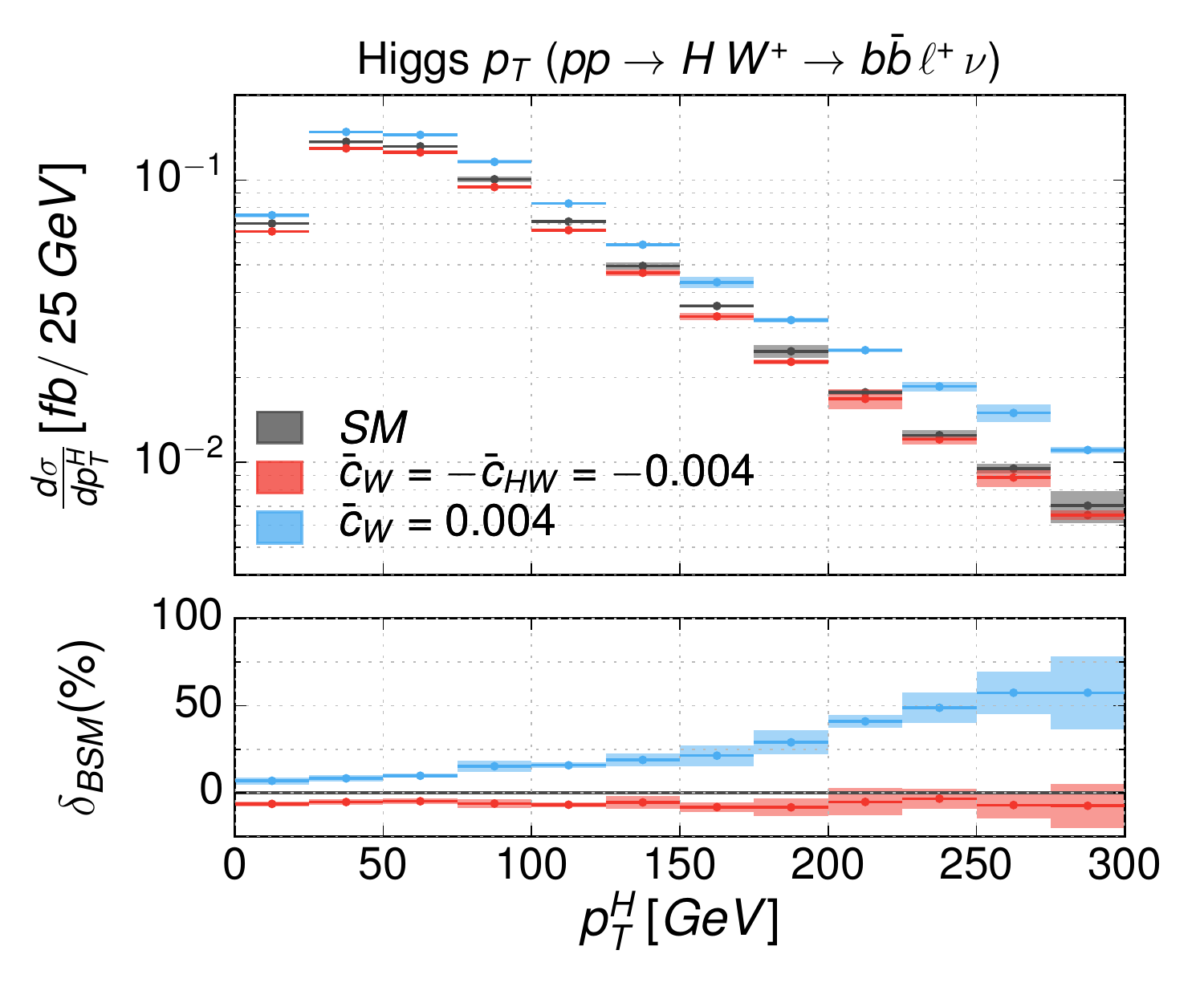}
\includegraphics[width=4.9cm]{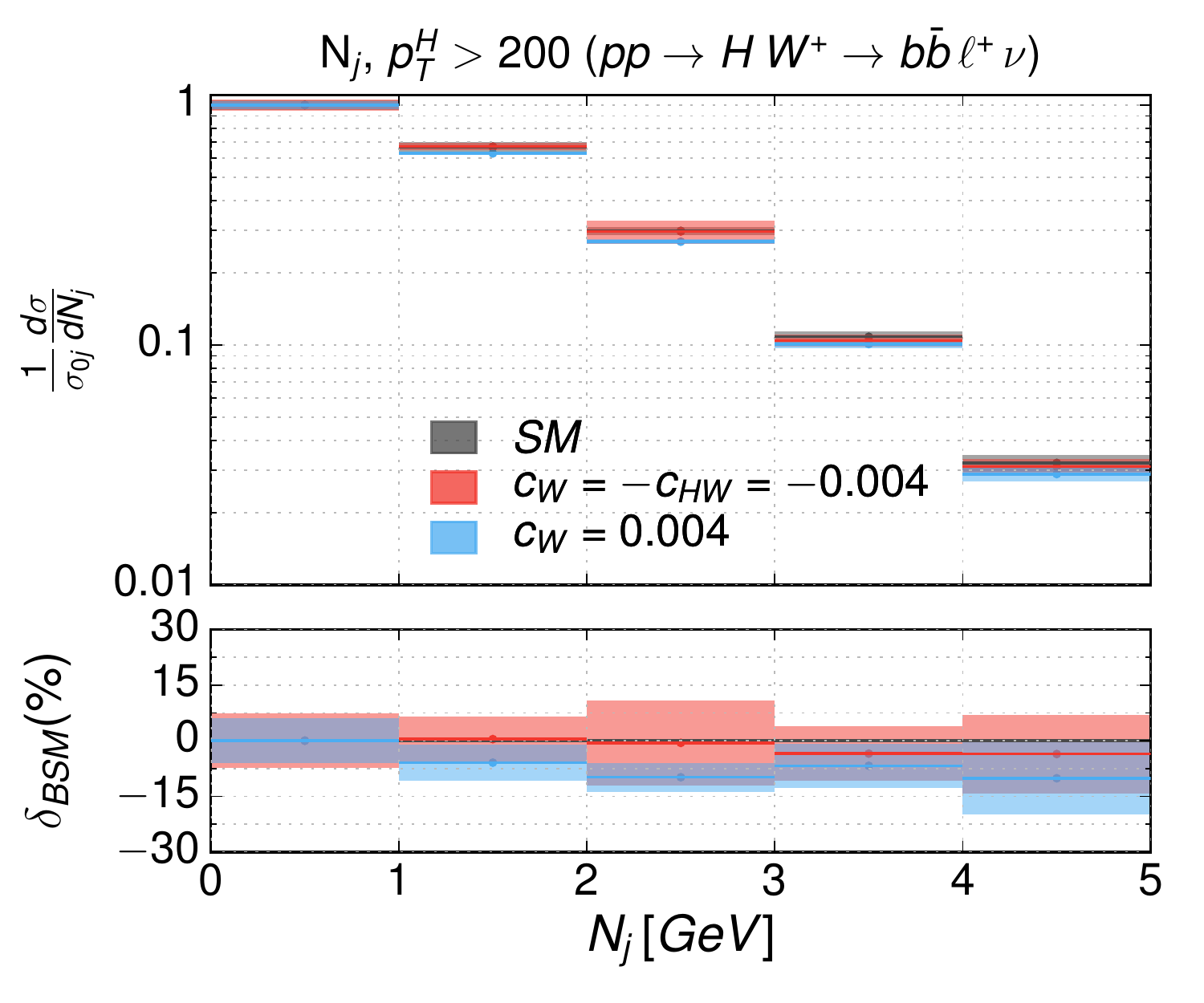}
\caption{\label{fig:VH_NLOPS} Comparison of differential distributions in the SM and the two EFT benchmarks of $\bar c_W=0.004$ and $\bar c_W=-\bar c_{HW}=-0.004$ using {\sc Powheg} + {\sc Pythia8}. Lower panels show the percentage deviation of the EFT benchmarks from the SM prediction, $\delta_{BSM}$.\\
\emph{upper row}: $pp\to Z H\to \ell^+\ell^- b\bar{b}$. From left to right-- the Higgs-Z invariant mass, $M_{VH}$; Z-boson transverse momentum, $p_T^Z$; and the number of jets normalised to the 0-jet bin, $N_j$.\\
\emph{lower row}: $pp\to W^+ H \to \ell^+ \nu b\bar{b}$. From left to right -- the transverse mass of the system, $m_T$ (defined in text);  Higgs transverse momentum, $p_{T}^H$; and the number of jets normalised to the 0-jet bin, $N_j$.}
\end{center} 
\end{figure}

For $ZH$ production, Figure~\ref{fig:VH_NLOPS} shows differential distributions with respect to the Higgs-Z invariant mass, Z-boson $p_T$ and the number of jets ($N_j$) normalised to the 0-jet bin comparing the SM to the two EFT benchmarks. For the $N_j$ distribution, an additional cut on the Higgs $p_T$ of 200 GeV is applied in order to isolate the region where the EFT contributions are most important. In the case of $WH$ production, the leptonic decay of the $W^+$ includes a neutrino which contributes to real missing energy on the event, preventing the construction of some of the kinematic variables available to the $Z$-boson associated production process, namely $m_{VH}$ and $p_T^V$, the total invariant mass and the vector boson transverse momentum. We trade these two observables for the total transverse mass of the system and the transverse momentum of the Higgs boson. Here, the total transverse mass of the $HW$ system is defined including the two $b$-jets, the lepton and the missing transverse energy,
\begin{align}
    m_T^2 = \left(\sum_i E^i_T + \slashed{E}_T\right)^2 - \left(\sum_i \vec{p}^{\,i}_T+ \slashed{\vec{p}}_T \right)^2;\quad i=b,\bar{b}, l^+.
\end{align}
The observable is the analogue of $M_{VH}$ in the $ZH$ case and is an approximation of the momentum flowing through the $WH$ vertex. These variables are shown in Figure~\ref{fig:VH_NLOPS} along with the normalised $N_j$ distributions, as in the $ZH$ case.

We see that these more conservative choices for the Wilson coefficients still permit $\mathcal{O}$(20--50\%) deviations in the tails of the various distributions for the $\bar c_W = 0.004$ benchmark with a clear preference for large momentum flow through the vertex. We also observe that the size of the deviation in the $M_{VH}$ distribution for $ZH$ correspond roughly to the size of the deviation in $p_T^Z$ at the corresponding energy scale, \emph{i.e}, $M_{VH}\sim p_T^Z/2$, demonstrating the expected correlation between the two observables. The second benchmark of $\bar c_W = - \bar c_{HW} = -0.004$ does not exhibit such large deviations, instead contributing a relatively flatter depletion of the differential rate. This can be traced to the different Lorentz structure governing the effective vertex. The difference between the two benchmarks shows that $g^{(1)}_{hvv}$ leads to much more striking `EFT-like' deviations than $g^{(1)}_{hvv}$. Looking more closely at the Feynman rules of Figure~\ref{f:feynrules}, we see that $g^{(2)}_{hvv}$ goes as the square of the individual momenta of the Z bosons, while $g^{(1)}_{hvv}$ goes as the product of the two Z boson momenta. As a consequence, in high centre of mass energy limit of the $ZH$ production matrix element, $g^{(2)}_{hvv}$ leads to a richer energy dependence, containing terms proportional to higher powers of Mandelstam variables $\propto st/M_Z^2,\,t^2/M_Z^2$ that are not present when only considering  $g^{(1)}_{hvv}$ contributions. The $N_j$ distributions -- although suffering from somewhat low MC statistics due to the $p_T^V>200$ GeV requirement -- appear to follow a similar trend.
\section{Conclusions}  \label{sec:concs}
 
Physics Beyond the Standard Model is likely to be connected to the Higgs sector, generically leading to deviations in the Higgs behaviour with respect to SM predictions. These indirect probes of new physics require a precise understanding of the SM contributions as well as the interplay between the SM and New Physics in observables. 
Among the different LHC Higgs observables, the production in association with a vector boson is specially sensitive to effects of new heavy particles in kinematic distributions and ratios of cross sections~\cite{withHwang,withJohn}.  

In this paper we have presented predictions for the associated production of a Higgs 
boson in association with a $W$ or $Z$ vector boson, including anomalous couplings 
between the Higgs and vector boson, not present in the Standard Model. Our predictions 
include effects in QCD beyond the Leading Order in perturbation theory. We presented predictions 
at fixed order (NLO) and matched to parton showers using the {\sc Powheg} formalism (NLO+PS). 

Anomalous couplings in the $HVV$ vertex (HAC) can arise in many extensions of the SM. 
A general model independent parameterization can be obtained by saturating 
the Lorentz structures of the four-dimensional $HVV$ interactions in the Lagrangian. Particular 
models then correspond to some (or all) of the new couplings acquiring non-zero values. 
An interesting class of models arise when the scale of new physics is large and can be integrated out 
of the Standard Model. In these scenarios the SM is treated as an effective field theory (EFT). We matched our 
results from the EFT to a linearly realized breaking of the EW symmetry, in which the Higgs is a doublet of $SU(2)_L$. Transitioning between 
the two calculations setups is straightforward, and we presented results in both the HAC and EFT frameworks. 

In order to maximize the physics potential of the LHC it is essential that precise theoretical predictions are used 
to compare theoretical predictions and experimental data. Matched parton showers, which combine the normalization 
and matrix elements of a Next-to-Leading Order calculation, and a leading logarithmic resummation of soft collinear logarithms 
provide a good framework for comparing theoretical predictions to data. The {\sc Powheg}-Box provides a public 
format to match results obtained at fixed order to parton showers, allowing for full event simulation. We calculated the NLO corrections 
to $VH$ production including the effects of anomalous couplings using analytic amplitudes and the spinor helicity formalism. 
We implemented this calculation into MCFM and modified the existing $VH$ processes in {\sc Powheg} to incorporate our new matrix elements. 

We used our results to study the phenomenological impact of our calculation at the Run II of the LHC operating at  $\sqrt{s}=13$ TeV. 
We demonstrated the capabilities of our code both at fixed order and NLO+PS accuracy generating events and showering them with the PYTHIA 
parton shower. We focused on parameter selections which are consistent with limits obtained during Run I. In this region NLO effects change 
the differential distributions by around $\mathcal{O}(20\%)$. 
Our results will be made publicly available in the released versions of the MCFM and {\sc Powheg} codes. 
\section*{Acknowledgements}

The work of KM and VS is supported by the Science Technology and Facilities Council (STFC) under grant number ST/J000477/1. 
\clearpage
\appendix
\section{Fields redefinitions and their contributions to EW parameters and gauge boson interactions\label{app:redefs}}
After electroweak symmetry breaking, the SM supplemented by the dimension-6 operators in eq.~\eqref{eq:SILH} leads to the following, non-canonical kinetic terms for the weak gauge bosons:
\begin{align}
    \nonumber
    \mathcal{L}_\mathrm{kin.} =&
    -\frac{1}{2}W^{\mu\nu}_+W_{\mu\nu}^-\Bigg[ 
       1 - \frac{v^2}{\Lambda^2}
       \frac{g^2}{4}\bar{c}_{\sss W}
       \Bigg]
     -\frac{1}{4}W^{\mu\nu}_3W_{\mu\nu}^3\Bigg[ 
        1 - \frac{v^2}{\Lambda^2}
        \frac{g^2}{4}\bar{c}_{\sss W}
        \Bigg]\\
    \nonumber
       &-\frac{1}{4}B^{\mu\nu}B_{\mu\nu}\Bigg[ 
       1 - \frac{v^2}{\Lambda^2}\frac{g^{\prime\,2}}{2}
       \left(\bar{c}_{\sss BB}+\bar{c}_{\sss B}\right)
       \Bigg]
       -B^{\mu\nu}W_{\mu\nu}^3\Bigg[\frac{v^2}{\Lambda^2}\frac{gg^\prime}{8}
       \left(\bar{c}_{\sss B}+\frac{\bar{c}_{\sss W}}{2}\right)
       \Bigg].
\end{align}
The following field redefinitions canonically normalise these terms and remove the $T_3$-Hypercharge mixing.
\begin{align}
    \nonumber
    W^\mu_{\pm} \to& 
    W^\mu_{\pm}\left[1+\delta W\right]\\\nonumber
    B^\mu \to& 
        B^\mu\left[1+\delta B\right] + y W^\mu_3\\
    W^\mu_3 \to& 
        W^\mu_3\left[1+\delta W\right] + z B^\mu\\
        \delta W=&\frac{v^2}{\Lambda^2}\frac{g^2 }{8} \bar{c}_{\sss W};\quad
    \delta B =\frac{v^2}{\Lambda^2}\frac{g^{\prime\,2}}{4}
       \left(\bar{c}_{\sss BB}+\bar{c}_{\sss B}\right)\\
        y+z \equiv&\kappa_{WB}= -\frac{v^2}{\Lambda^2}\frac{gg^\prime}{4}
       \left(\bar{c}_{\sss B}+\frac{\bar{c}_{\sss W}}{2}\right)
\end{align}
One may also redefine the weak and hypercharge gauge couplings in order to absorb the effects of some of these shifts.
\begin{align}
    g\to\bar{g}\left[1+\delta g\right];\quad
    g^\prime\to\bar{g}^\prime\left[1+\delta g^\prime\right]
\end{align}
In general, the $W$ mass is modified and the neutral mass matrix has one zero eigenvalue for the photon and a modified mass for the $Z$ boson. The following choice for the gauge coupling redefinitions preserves the SM expressions for $W$-boson mass and interactions as well as the definition of the Weinberg angle in terms of the gauge couplings while shifting the $Z$-mass:
\begin{align}
    \delta g &=-\delta W ;\quad \delta g^\prime = - \delta B + \delta W+\delta g,
    -\frac{\bar{g}^\prime}{\bar{g}}y+\frac{\bar{g}}{\bar{g}^\prime}z\\
    \Rightarrow m_{W}&=\frac{e v}{2\hat{s}_W};\quad
    m_{Z}=\frac{e v}{2\hat{s}_W\hat{c}_W}\left[
    1+\delta m_{Z}
    \right],
\end{align}
with
\begin{align}
    \hat{c}_{W} =&\frac{\bar{g}}{\sqrt{\bar{g}^2+\bar{g}^{\prime\,2}}};\quad
    e=\bar{g}\hat{s}_W;\quad \delta m_{Z}=\frac{\hat{s}_W}{\hat{c}_W}(z-2\kappa_{WB}).
\end{align}
We choose to set the parameter $z$ to:
\begin{align}
    z=&-\frac{e^2v^2}{8\Lambda^2}\frac{\hat{s}_W}{\hat{c}_W}\bar{c}_W.
\end{align}
The shifts to the fermionic photon and $Z$ couplings parametrised as:
\begin{align}
    Q^\prime =& eQ\left[ 1+ \delta e\right],\\
    Q^\prime_Z =& \frac{e}{\hat{c}_W\hat{s}_W}\left[
    T_3\left(1 + \delta T^Z_{3}\right)
    -Q\hat{s}^2_W\left(1 + \delta Q^Z\right)\right],
\end{align}
are given in eqs.~\eqref{eq:ashift} and ~\eqref{eq:zshift}. We note that the difference between using the SM and EFT definitions of parameters multiplying a dimension-6 $\left(\mathcal{O}(\Lambda^{-2})\right)$ contribution is higher order in the EFT expansion.

The extraction of the EW parameters from the $\{m_{Z},m_{W},G_F\}$ set of inputs discussed in sec.~\ref{sec:implementation} follows from these definitions. It is important to stress that these definitions are valid for the subset of operators that are implemented in our code, namely $\mathcal{O}_W,\mathcal{O}_{HW},\mathcal{O}_B,\mathcal{O}_{HB}$ and $\mathcal{O}_\gamma$. In general, the presence of a complete dimension-6 basis of operators will lead to more modifications, such as with the $\mathcal{O}_H$ and $\mathcal{O}_T$ operators modifying the canonical normalisation of the Higgs field and therefore its couplings. A consequence of this can be seen in Table~\ref{trans}, where these two Wilson coefficients appear in the SM-like $g^{(3)}_{hzz}$ structure.

\end{document}